\documentclass{cjpsuppl}
\usepackage{graphics} 
\usepackage{epsfig}
\usepackage{wrapfig}

\newcommand{\itemnew}{\item \vspace*{-2mm}}
\def\etal{{\sl et al.}} 
\newcommand{\rar}{\rightarrow}
\newcommand{\pdup}{p_\uparrow}

\newcommand{\upr}{\uparrow}
\newcommand{\pup}{p^\uparrow}

\newcommand{\ddup}{d_\uparrow}
\newcommand{\pimp}{\pi^- + \pdup \rar \pi^0 + X}

\newcommand{\pimall}{\pi^- + \pdup (\ddup) \rar \pi^0 (\eta) + X}

\newcommand{\PRD}[1]{{\it Phys.\ Rev.}\ {\bf D#1}}
\newcommand{\PRL}[1]{{\it Phys.\ Rev.\ Lett.}\ {\bf #1}}
\newcommand{\PLB}[1]{{\it Phys.\ Lett.}\ {\bf B#1}}

\newcommand{\PAN}[1]{{\it Phys.\ Atom.\ Nucl.}\ {\bf #1}}
\newcommand{\YAF}[1]{{\it Yad.\ Phys.}\ {\bf #1}}
\newcommand{\NIMs}[1]{{\it NIM}\ {\bf #1}}
\newcommand{\xf}{x_{\mathrm F}}
\newcommand{\ppdup}{p + \pdup \rar \pi^0 + X}

\begin{document}
\slacs{.6mm}
\title{Spin physics with light and heavy neutral mesons at Protvino}
\authori{V.V.~Mochalov, S.V.~Ivanov, V.I.~Garkusha, A.S.~Gurevich, 
V.I.~Kravtsov, O.P.~Lebedev, N.I.~Minaev, L.V.~Nogach, S.B.~Nurushev, 
A.N.~Vasiliev}
\addressi{IHEP, Protvino, Russia} 
\authorii{A.V.~Otboev, Yu.M.~Shatunov, D.K.~Toporkov}    
\addressii{BINP, Novosibirsk, Russia}
\authoriii{A.S.~Belov}
\addressiii{INR, Moscow, Russia}
\authoriv{}    \addressiv{}
\authorv{}     \addressv{}
\authorvi{}    \addressvi{}
\headtitle{Spin physics with light and heavy neutral mesons 
at Protvino}
\headauthor{V.~Mochalov}
\lastevenhead{V.~Mochalov: Spin physics with light and heavy 
neutral mesons at Protvino}
\pacs{13.20.Cz, 13.20.Fc, 13.20.Gd, 13.88+e}
\keywords{Nucleon spin, polarization, transversity, single-spin 
asymmetry, double-spin asymmetry, charm production}
\refnum{}
\daterec{} 
\suppl{?}  \year{2006} \setcounter{page}{1}
\maketitle

\begin{abstract}
PROZA-M experiment results as well the proposal of a new spin 
program with the use of a polarized proton beam are presented. 
Significant asymmetries were observed in inclusive $\pi^0$ 
production. 
The new program proposes to study a wealth of single- and 
double-spin observables in various reactions using 
longitudinally and transversely polarized 
proton beams at U70. The main goal is to define gluon 
contribution to nucleon spin by measuring double-spin 
asymmetry in charmonium production. 
\end{abstract}

\section{Introduction} 

Polarization experiments give us an unique opportunity to probe the 
nucleon internal structure. While spin averaged cross-sections 
can be calculated within acceptable accuracy, current theory of 
strong interactions can not describe large single-spin asymmetries 
and polarization. Unexpected large values of  single spin asymmetry (SSA) 
in inclusive $\pi$-meson production are real challenge 
to current theory because naive perturbative Quantum Chromodynamics (QCD) 
predicts small asymmetries due to the helicity conservation
decreasing with transverse momentum increase.

Although years of experimental efforts at miscellaneous accelerators
have provided a lot of information about the QCD hard scattering
and the parton structure of the proton, there is no corresponding 
body of data on the spin-dependence of the elementary interactions 
and the spin structure of the proton. High intensity polarized 
proton beam to be accelerated up to 70~GeV and extracted from U70 
main ring would offer the opportunity to study the unique properties 
of the spin variable at large $x$ to increase the
understanding of these fundamental quantities.

We present the result of SSA measurements at PROZA experimental setup
(section \ref{proza-result}) as well as the proposal of new experiments
with polarized proton beam at 70 GeV proton synchrotron of IHEP-Protvino
(section \ref{pol-beam}).

\section{Single-spin asymmetry in inclusive $\pi^0$ production.}
\label{proza-result}

The PROZA collaboration started to study SSA measurements of neutral 
mesons 30 years ago. Comprehensive study of exclusive 
charge-exchange reactions pointed out to significant spin 
effects~\cite{proza-exclusive}. To investigate the dependence of the
asymmetry on beam particle flavor and energy as well as on secondary 
meson production angle, the measurements were carried out at the central 
and polarized target fragmentation region at the energy range 40-70 GeV 
in the reactions:
\bea
\pimall 
\label{eq:pimpi0}\\
\ppdup
\label{eq:pppi0} 
\eea

Single spin asymmetry $A_N$ is defined as
\be
A_N(x_F,p_T)= \frac{1}{P_{targ}}\cdot 
\frac{1}{<cos\phi>}\cdot 
\frac{\sigma_{\uparrow}^H(x_F,p_T)-\sigma^H_{\downarrow}(x_F,p_T)}
{\sigma^H_{\uparrow}(x_F,p_T)+\sigma^H_{\downarrow}(x_F,p_T)}
\label{eq:asymdef}
\ee
\noindent
where $P_{targ}$ is the target polarization,  $\phi$ is the 
azimuthal angle between the target-polarization vector
and the normal to the plane spanned by the beam axis and the momentum 
of the outgoing neutral pion, and $d\sigma^H_{\uparrow}$ 
($d\sigma^H_{\downarrow}$) are the invariant differential cross
sections for neutral-pion production on hydrogen for
opposite directions of the target-polarization vector.
We detected neutral pions in the azimuthal 
angle range of $180 \pm 15 ^{\circ}$; therefore, we set $\cos \phi= -1$.  
Since the $\pi^0$'s detection efficiency is identical for the two 
directions of the target polarization vector, we find for 
the detector on the right side of the beam, that
\be
A_N=-\frac{D}{P_{targ}}\cdot A_N^{raw} =
-\frac{D}{P_{target}}\cdot
\frac{n_{\uparrow}-n_{\downarrow}}{n_{\uparrow}+n_{\downarrow}}
\label{eq:asymreal}
\ee 
\noindent
where $A_N^{raw}$ is the raw asymmetry actually measured
in the experiment, $D$ is the target-dilution factor, and
$n_{\uparrow}$ ($n_{\downarrow}$) are the normalized 
(to the monitor) numbers of detected neutral pions for up and down 
directions of the target-polarization vector. The procedure used
to calculate $D$ was described in detail elsewhere \cite{protv_yaf}. 

\begin{figure}[b]
\centering
\includegraphics[width=0.8\textwidth]
{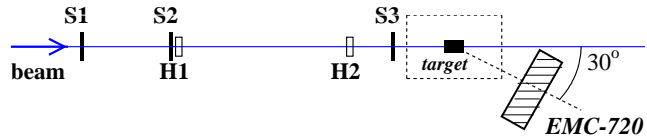}
\vspace*{-1.3cm}
\caption{ Experimental Setup PROZA-M. S1-S3 -- trigger scintillation 
counters; H1-H2 -- hodoscopes; $EMC1-EMC3$ -- electromagnetic calorimeters;
$target$ -- polarized target.}
\label{fig:setup}
\end{figure}

Current configuration of the PROZA-M Experimental Setup \cite{setup} is 
presented on {\bf Fig.~\ref{fig:setup}}. The beam of negatively charged 
particles produced in the internal target was deflected into the beam-line 
by guide magnetic field of the accelerator. For the first time the proton 
beam was extracted from a strong-focusing synchrotron U70 by the thin
$Si$ crystal bent under 80 mrad \cite{Aseev,beam}. A frozen 
propane-diol ($C_3H_8O_2$) target with an average polarization 85\% was 
used. \cite{target}. Beam particles were detected by the two 
two-coordinate hodoscopes $H1-H2$, placed 8.7 and 3.2~m downstream 
the target.  Three scintillation counters $S1-S3$ 
were used for a zero level trigger. $\gamma$-quanta were detected by 
the total absorption electromagnetic calorimeters $EMC$ 
(array of 720 lead-glass cells \cite{steklo}). The  counters were of size 
$38 \times 38 \times 450$~mm$^3$ (18~rad. length). Currently the 
detector is situated 2.3 m downstream the target at 
$30^{\circ}$  at laboratory frame to measure the asymmetry in the 
polarized target fragmentation region.

\begin{figure}[t]
\centering
\begin{tabular}{cc}
\includegraphics[width=0.45\textwidth,height=3.7cm]
{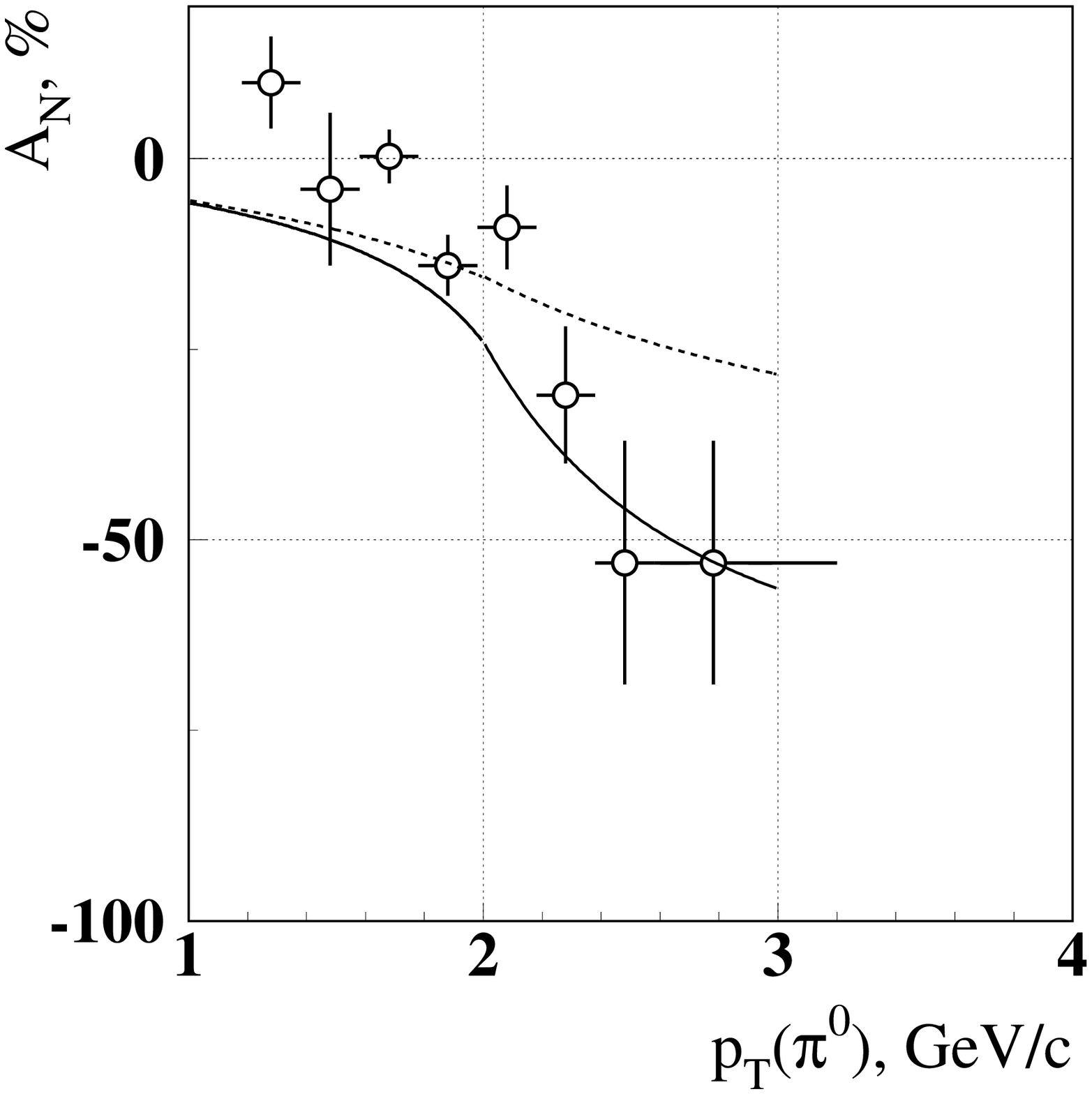} &
\includegraphics[width=0.45\textwidth,height=3.7cm]
{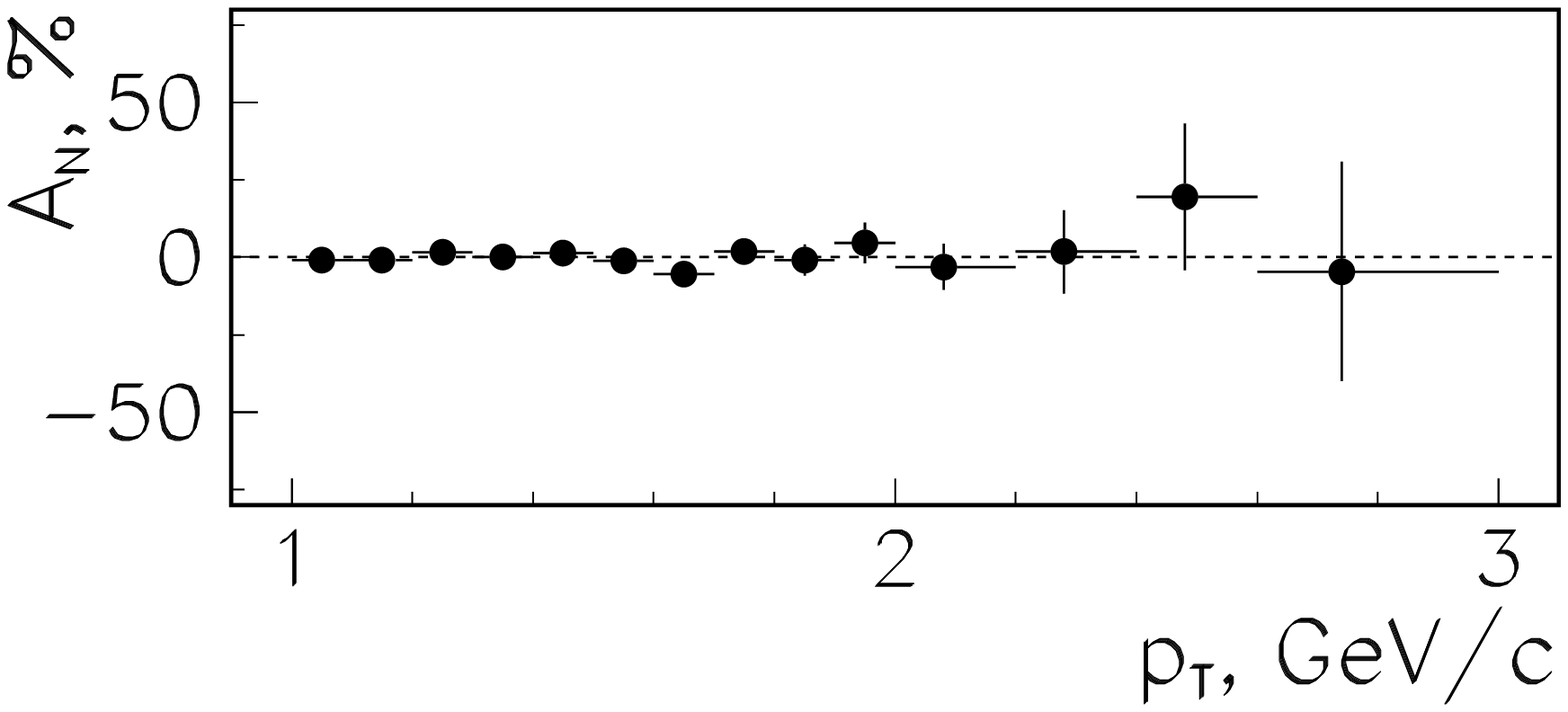} 
\\
\end{tabular}
\vspace*{-0.8cm}
\caption{Left: summary $A_N$ for the reaction $\pimall$; 
right: $A_N$ for the reaction $\ppdup$; both at the central 
region. Solid line -- predictions for U-matrix 
model \cite{troshin40}, dashed -- for the model with quark 
chromomagnetic moment \cite{ryskin}.}
\label{fig:sumcentr}
\vspace*{-0.4cm}
\end{figure}

$A_N$ in the central region was measured for the reactions
$\pimall$ \cite{protv_yaf,protv_plb} and in the reaction
$\ppdup$ \cite{protv70cent} in wide range of transverse
momenta ({\bf Fig.~\ref{fig:sumcentr}}).
Asymmetry is consistent with zero in $pp$-interaction, while
significant effect was observed in the reaction (\ref{eq:pimpi0}). 
One may conclude that the asymmetry in the central region depends
on quark flavor.

\begin{figure}[b]
\centering
\begin{tabular}{cc}
\includegraphics[width=0.45\textwidth,height=3.5cm]
{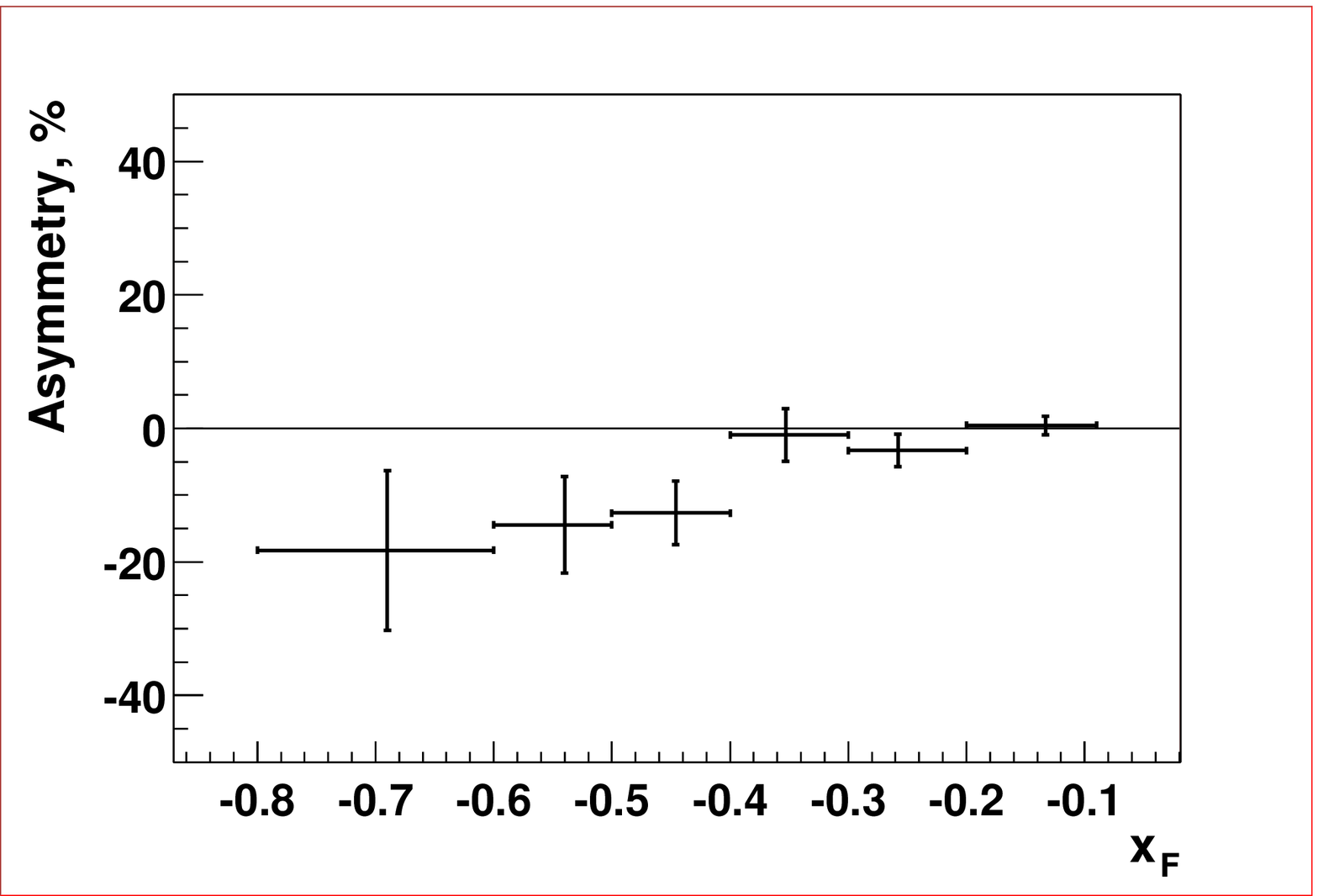} 
&
\includegraphics[width=0.45\textwidth,height=3.5cm]
{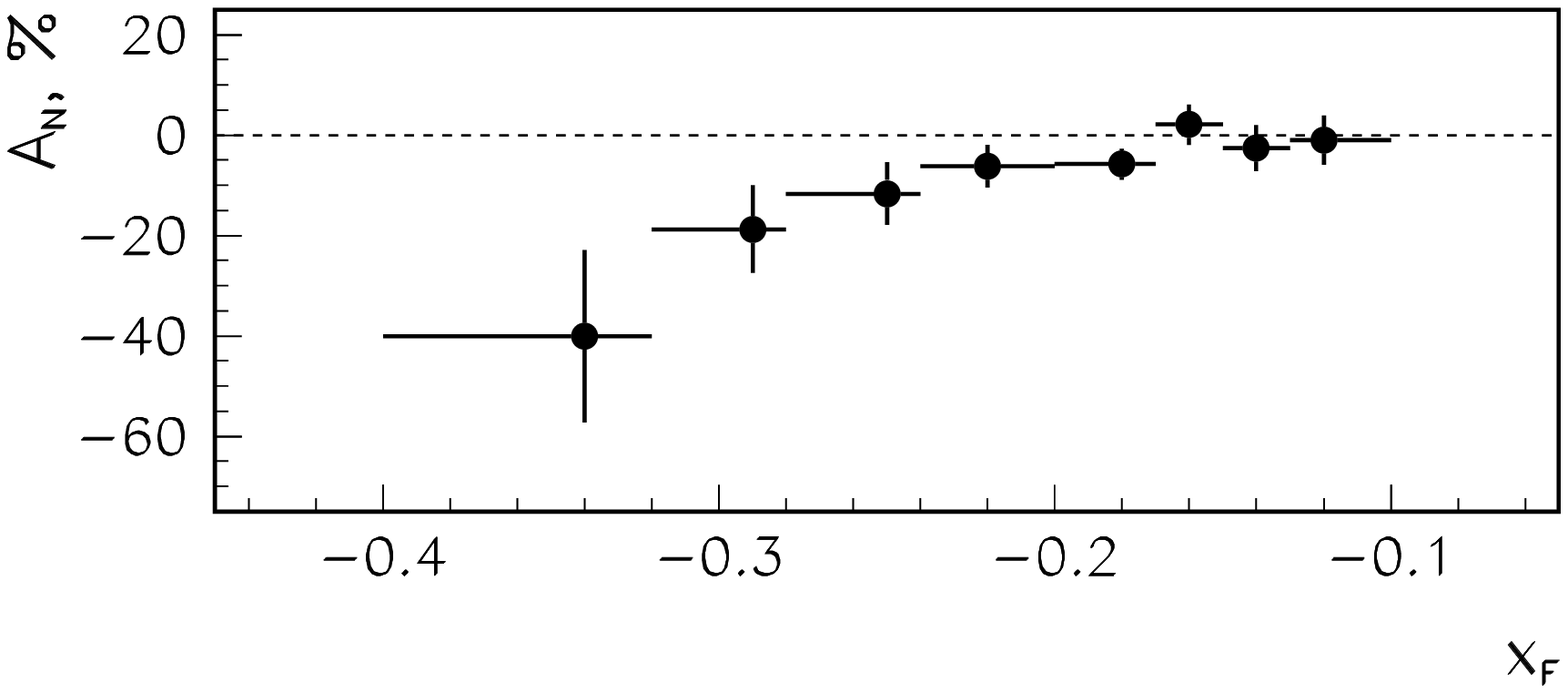} 
\\
\end{tabular}
\vspace*{-0.8cm}
\caption{$A_N$ in the reaction $\pimp$ at 40 GeV (left) and 
in the reaction $\ppdup$ at 70 GeV (right) in the 
polarized target fragmentation region.}
\label{fig:sumback}
\end{figure}  

SSA in the target fragmentation region (see {\bf Fig.~\ref{fig:sumback}}) 
was measured in the reaction $\ppdup$ at 70 GeV \cite{protv70back} and 
in the reaction $\pimp$ at 40 GeV \cite{protv40back}. 
Asymmetry is consistent with zero at small 
absolute values of scaling variable $\xf$, then starts 
to grow up in the magnitude and achieves significant values.
The asymmetry does not depend on beam particle flavor in the 
polarized target fragmentation region. 

\begin{wrapfigure}{R}{7.cm}
\vspace*{-1.cm}
{\hspace*{.3cm}
\mbox{\epsfig{figure=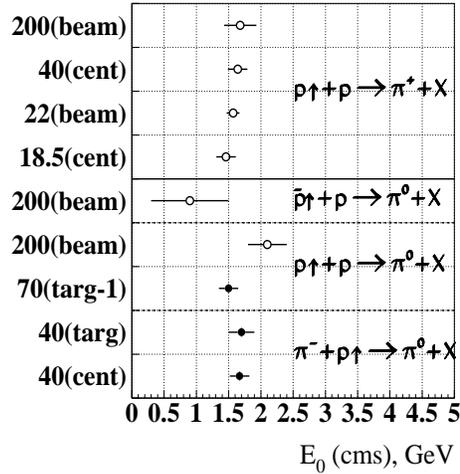,
width=6.6cm,height=7.cm}}} \\
\hspace*{0.3cm}
\begin{minipage}[t]{6.6cm}
\vspace*{-0.8cm}
\caption{{Center of mass energy values where  the pion asymmetry 
starts to grow up for different experiments.
The energy along the Y-axis is in GeV; $cent$ -- corresponds to 
experiments in the central region ( $x_f\approx0$), 
$targ$ -- the polarized target fragmentation region; 
$beam$ -- the polarized beam fragmentation region.}
}
\label{fig:threshold}
\end{minipage}
\vspace*{-0.4cm}
\end{wrapfigure}

Universal threshold of single-spin asymmetry in inclusive 
pion production was found for fixed target experiments 
\cite{threshold}. The asymmetry is consistent with zero
below threshold energy value in the center of mass system and
then starts to grow up. The result of the threshold 
energy for different experiments is presented in 
{\bf Fig.~\ref{fig:threshold}}. The effect may 
points out the existence of complex structure inside 
constituent quark \cite{const_quark}. 

We continue to study asymmetry in the reaction $\ppdup$ at 
intermediate and large negative values of $\xf$. 
During first stage of the measurements we have accumulated enough 
statistics to achieve accuracy in $pp$-interaction at 
intermediate negative values of $\xf$ at the same or even better 
level as for $\pi^-p$-interactions. The expected accuracy
will allow us to check ``universal threshold'' . 
The statistics in the reaction(\ref{eq:pppi0}) will be increased 
at least by factor of four during 2007 data-taking run, 
especially for large negative values of $\xf$. 

Nevertheless to investigate universal threshold for different
particles and in different kinematic regions polarized proton 
beam is required. 

\section{Experiments with polarized proton beam at U-70}
\label{pol-beam}

New program of spin experiments is developing now at IHEP. 
We propose to produce a polarized proton beam from the polarized
atomic beam-type source, accelerate it up to 70~GeV and
deliver to several experimental setups to:
\begin{itemize}
\itemnew  measure the gluon and quark polarization in longitudinally 
polarized protons in charmonium production;
\itemnew study transversity distribution by measuring the 
double-spin asymmetry for the Drell-Yan muon pair production;
\itemnew measure the dependence of single-spin asymmetries on
separate kinematic variables $p_T$ and $\xf$ and hadron flavor 
to study the it's origin;
\itemnew measure miscellaneous spin parameters in hyperon
production at moderate transverse momenta to learn about the 
role of strange quarks in the spin structure of nucleon;
\itemnew measure polarization and spin correlation parameters in elastic
$pp$-scattering in the hard scattering region in order to check the
QCD predictions.
\end{itemize}

Final goals of spin physics at U70 are :
\begin{itemize}
\itemnew
study the spin structure of the proton, i.e., how the proton's 
spin state can be obtained from a superposition of Fock states 
with different numbers of constituents with nonzero spin;
\itemnew
study how the dynamics of constituent interactions depends 
on  spin degrees of freedom and on the flavors;
\itemnew
understand chiral symmetry breaking and helicity 
non-conservation on the quark and hadron levels;
\itemnew study the overall nucleon spin structure in the 
range of moderate $p_T$(up to 5 GeV/c), and the long range 
QCD dynamics (confinement), including study non-perturbative 
interactions of massive constituent quarks with an effective 
color field of flux tubes, produced by confined
quarks and gluons.
\end{itemize}

These issues are closely interrelated at the hadron level and
the results of the experimental measurements
are to be interpreted in terms of hadron  spin
structure convoluted with the constituent interaction dynamics.

\subsection{Acceleration of polarized proton beam}
To do this spin physics, the polarized proton beam with an 
intensity up to  $5 \cdot 10^{12}$ protons per spill and energy 
up to 70 GeV needs to be accelerated. 
This process requires the development of intensive polarized proton 
source, acceleration of the polarized beam in 1.5~GeV booster
and U-70 main ring, extraction of the beam onto experimental setup.
Development of beam polarimetry and polarized targets is also required

\subsubsection{Polarized Ion Source}

To reach an intensity of the polarized proton beam in the U-70 
accelerator higher than $5 \times 10^{12}$ p/cycle one needs to design 
and build a high intensity source. 
The technique of sources with optical pumping (optical) 
or with resonant charge-exchange plasma ionizer (atomic) 
has been successfully developed for last decades. The following characteristics of 
the existing and developing sources are presented 
in {\bf Table.~\ref{tab:source}}: type of the source, peak intensity $I$,  
polarization $P$, emittance, pulse duration,  repetition rate, figure of 
Merit  $P^2 \cdot I$ 
and modified figure of merit $P^2\cdot I\cdot N$ \footnote 
{$N$ is number of turns of polarized ion beam during its stripping injection 
into the booster ring.}.
\begin{table} [t]
\vspace*{-0.6cm}
{\small
\caption{Characteristics of different polarized ion sources}
\label{tab:source}
\begin{tabular}{|c|c|c|c|c|c|c|c|c|}
\hline
Lab. & Type         & $I$,  & P     & Emitt.      & Pulse & Rep., & $P^2\cdot I$ &$P^2\cdot I\cdot N$ \\
     &              & (mA)  &       & {\small (mm$\cdot$mrad)} & duration & rate    &        & \\
     &              &       &       &                          & (msec)   & (Hz)    &        & \\
\hline
\hline
RHIC & Optical      &      &       &                 &       &        &        &                 \\
(BNL) & $H^-_{\upr}$ & 1    &  0.82 & $2 \pi$    & 0.5   & 1      & 0.67   & 14.7 \\
\hline
{\small TRIUMF} &  Optical &      &       &                 &       &        &        &      \\
/INR   &  $H^-_{\upr}$ &  8   & 0.42  & $2 \pi$    & 0.1   & 1      & 1.4    & 30.8 \\
      &  $H^+_{\upr}$ & 50   & 0.5   & $2 \pi$    & 0.1   & 1      & 12.5   & 12.5 \\
\hline
INR,   &  Atomic     &      &       &                 &       &        &        &      \\
{\small Moscow}&  $H^-_{\upr}$ & 3.8 & 0.91  & $1.7 \pi$    & 0.17  & 5      & 3.15  & 69.3 \\
      &  $H^+_{\upr}$ & 11   & 0.8   & $1.0 \pi$    & 0.2   & 5      & 7.0   & 7.0 \\
\hline
IUCF  &  Atomic     &      &       &                 &       &        &        &      \\
      &  $H^-_{\upr}$ & 1.6 & 0.85  & $1.2 \pi$    & 0.3  & 2      & 1.2  & 26.4 \\
      &  $D^-_{\upr}$ & 1.8   & 0.9   & $1.2 \pi$    & 0.3  & 2      & 1.2  & 26.4 \\
\hline
COSY  &  Atomic     &      &       &                 &       &        &        &      \\
      &  $H^-_{\upr}$ & 0.02 & 0.9   & $0.5 \pi$  & 10  & 0.5   & 0.016  & 0.35 \\
      &  $D^-_{\upr}$ & 0.02 & 0.9   & $0.5 \pi$  & 10  & 0.5   & 0.016  & 0.35 \\
\hline
JINR  &  Atomic     &      &       &                 &       &        &        &      \\
      &  $D^+_{\upr}$ & 0.4 & 0.6   &             & 0.4  & 0.1   & 0.144  & 0.144 \\
\hline
\end{tabular}}
\vspace*{-.4cm}
\end{table}

A high intensity pulsed source of polarized protons and negative 
polarized hydrogen ions built at the Institute of Nuclear Research (INR) 
in Troitsk \cite{source} have the record figure of merit $P^2 \cdot I$. 
The polarized ions are being generated as a result of 
the charge exchange reaction between the polarized 
hydrogen atoms and the unpolarized deuterium ions 
in the deuterium plasma. This source gives polarized 
protons of 11 mA in the pulsed mode with polarization $P= 80$\%. 
For the negative polarized hydrogen atoms the pulsed current $I$ 
is 3.8 mA and the polarization is 90\%.  The similar source 
produced at IUCF in collaboration with INR was successfully used at the 
Indiana University in USA. The source operation demonstrated  a long 
term stability and reliability. The source current can be increased 
further by using of sextupole separating magnets with magnetic 
field up to 5 T. 

\smallskip
\subsubsection{Conservation of the polarization during 
acceleration}

During acceleration, the polarization may be lost when 
the spin precession frequency passes through a so-called 
depolarizing resonance. These resonances occur when the 
spin tune $\nu_{sp}=\gamma \times G$ (where $G=1.793$ is the 
anomalous magnetic moment of the proton and $\gamma=E/m$) is equal to 
an integer number (imperfection resonances), or equal to 
$kP\pm$$\nu_z$ (intrinsic resonances). 
Here P=12 is the superperiodicity of the U70 accelerator 
(or its booster), $\nu_z$ is the vertical betatron tune, and $k$ 
is an integer. Imperfection resonances are due to vertical 
closed orbit errors and intrinsic resonances are due to 
the vertical betatron motion.

There are four strong resonances in 1.5~GeV booster 
({\bf Table.~\ref{tab:res_booster}}). There is no space inside 
booster for pulsed quads to pass resonances
using fast jump of the betatron frequency. Nevertheless the polarization in 
booster is still expected to be conserved at the level of 70\% even without
any modification of the booster. We study three 
possible ways to decrease depolarization inside booster:
\begin{itemize}
\itemnew to inject the beam to main ring at the energy close to 
1~GeV. In this case last two resonances are obsolete
and the depolarization in booster will be 15\%
\itemnew the strength of imperfection resonances can be 
decreased by proper vertical orbit correction and decreasing of the beam 
emittance.
\itemnew the strength of last two resonances may be increased
to flip spin direction by some blow-up of beam emittance.
\end{itemize}

\begin{table}[t]
\centering
\vspace*{-.8cm}
\caption{Strong resonances at booster}
\label{tab:res_booster}
\begin{tabular}{|c|c|c|c|}
\hline
Resonance $\nu_z$ & $\nu_{sp} $ & $W_{kinet} $ & $P_f/P_i$ \\
\hline
\hline
2 & 2 & 0,108 & 0,96 \\
3 & 3 & 0,631 & 0,89 \\
4 & 4 & 1,155 & $-0,89$ \\
\hline
$0 + Q_Z$ &3,78 & 1,040 & $-0,93$ \\
\hline
\end{tabular}
\end{table}

\subsubsection{Polarization at main ring}

\begin{figure}[b]
\parbox{0.47\hsize}{
\includegraphics[width=\hsize]
{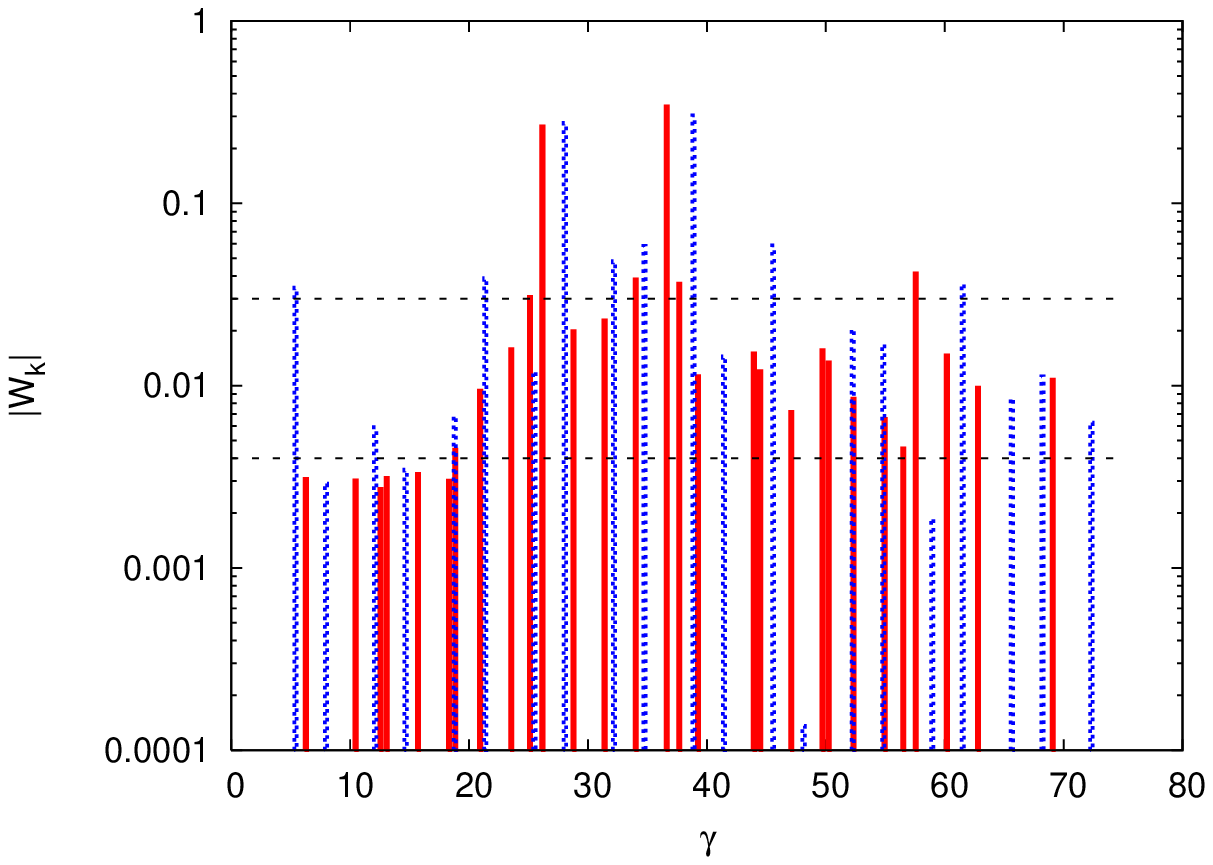} 
\caption{Resonances at U-70 main ring: solid -- imperfection, 
dashed -- intrinsic.}
\label{fig:resonances}}
\parbox{0.50\hsize}{
\includegraphics[width=\hsize]
{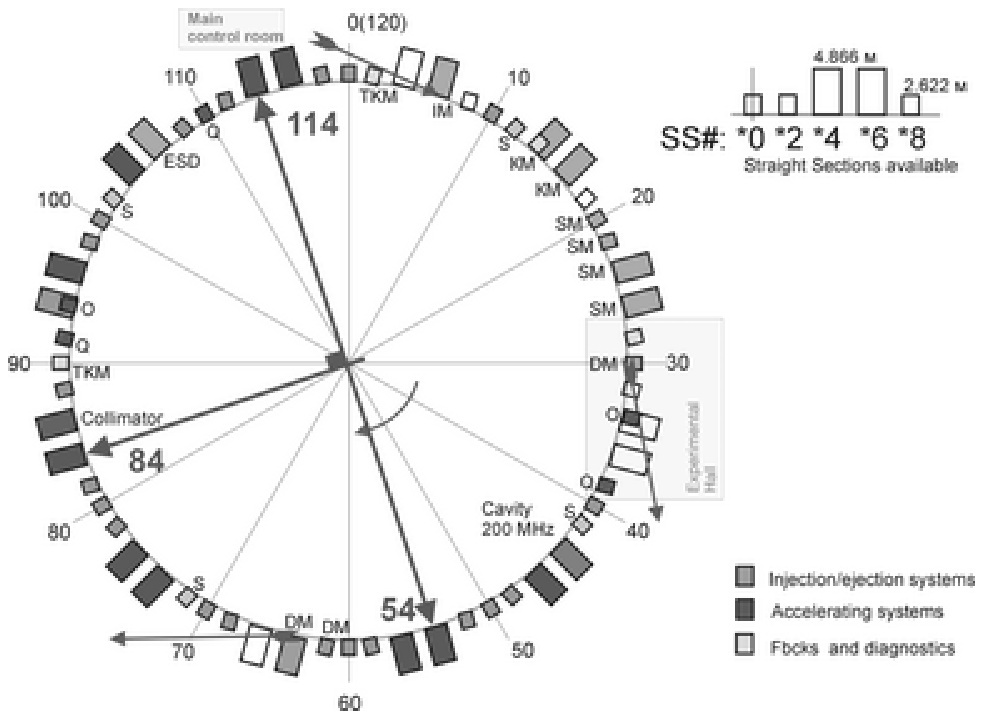}
\caption{Snake position proposed for the U-70.}
\label{fig:snake_u70}}
\end{figure}

The number of resonances at U-70 and 
their strength ({\bf Fig.~\ref{fig:resonances}}) is too 
large to overcome them without inserting into machine 
lattice a special array of magnets, which will rotate spin 
(syberian snakes) \cite{snake_derbenev}.  
There are a number of schemes with combinations of 
solenoids and skew quads, that don't cause the coupling 
out side the insertion. But such schemes require, 
as a rule, a long straight section. Compact partial 
snakes can be designed of helical magnets \cite{snake_rhic}.  
Spin rotations by transverse fields don't depend on energy 
for high energy particles whereas the action on spin by 
longitudinal fields is inversely proportional to the energy. 
Orbit distortions due to transverse fields can be minimized 
in the case of helical magnets by a dedicated snake design comprising a 
set of four full twist helixes with mirror symmetry and 
adjusted field levels.  A partial snake rotates spin only 
by a few degrees.  First beam test of this concept have been done in 
1976 year at electron-positron collider VEPP-2M \cite{snake_vepp}.
Later the solenoidal partial snake was applied at AGS, 
where polarized protons are accelerated up to 25 GeV.

\begin{figure}[t]
\centering
\begin{tabular}{cc}
\includegraphics[width=0.46\textwidth] 
{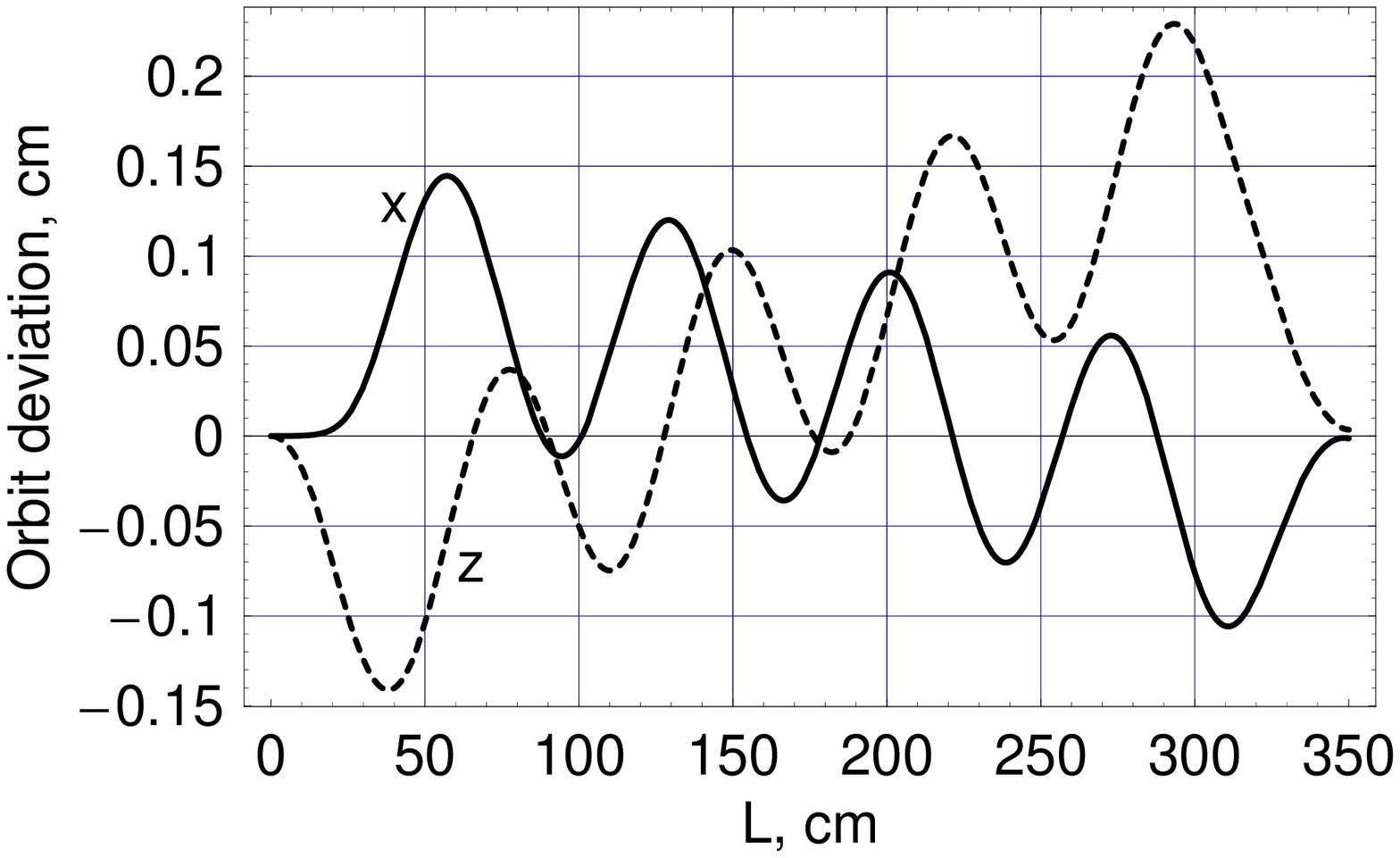} &
\includegraphics[width=0.46\textwidth] 
{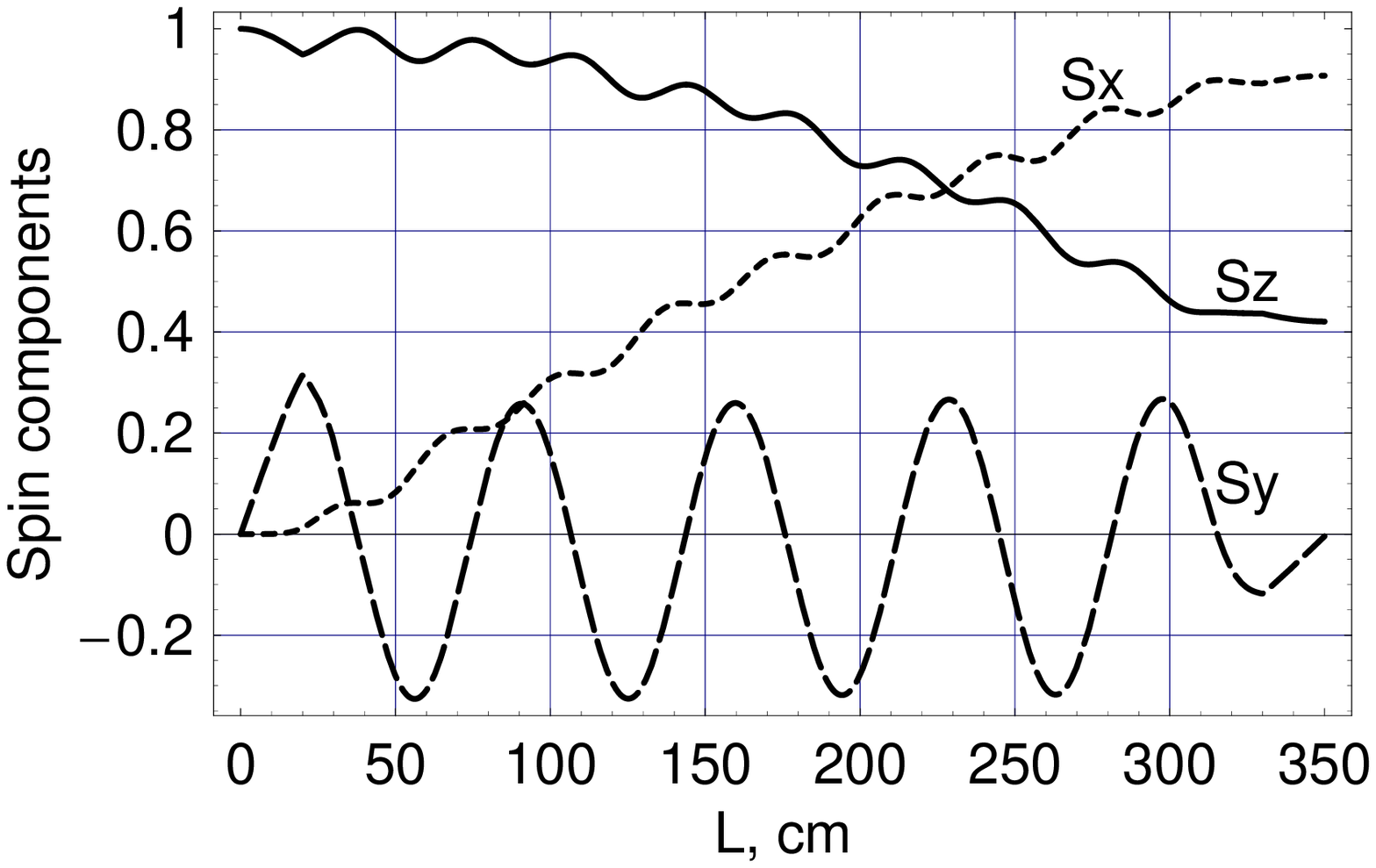} \\
\end{tabular}
\vspace*{-0.7cm}
\caption{Closed orbit distorsion (left) and spin rotation (right) 
inside the snake}
\label{fig:spin}
\vspace*{-0.4cm}
\end{figure}

The present proposal suggests installing three identical 
partial snakes into three super-periods of U-70, as it's 
shown in {\bf Fig.~\ref{fig:snake_u70}}. 
Each snake occupies 3.5 m longitudinally and consists of a 
4-period helix ($\lambda =70$~ cm; aperture 15~cm) with field up 
to 4.2~T. This combination rotates spin by 65$^{\circ}$. 
To compensate for some optics distortion ($<5$\%) two dipole 
coils are wounded at the edges. 

Particle trajectories inside the snake and spin rotation 
by the snake per pass are shown in {\bf Fig.~\ref{fig:spin}} 
for the beam energy 25 GeV.  The spin rotations by all the  
snakes with identical polarities are added coherently at 
$\nu_0 = k = m \cdot P/4$ ($m$ are integers) or at 
$\nu_0 = k = (m+1/2) \cdot P/4$ with one snake polarity being reversed. 
The last option looks more suitable for U-70. 

\begin{figure}[b]
\centering
\begin{tabular}{cc}
\includegraphics[width=0.49\textwidth] 
{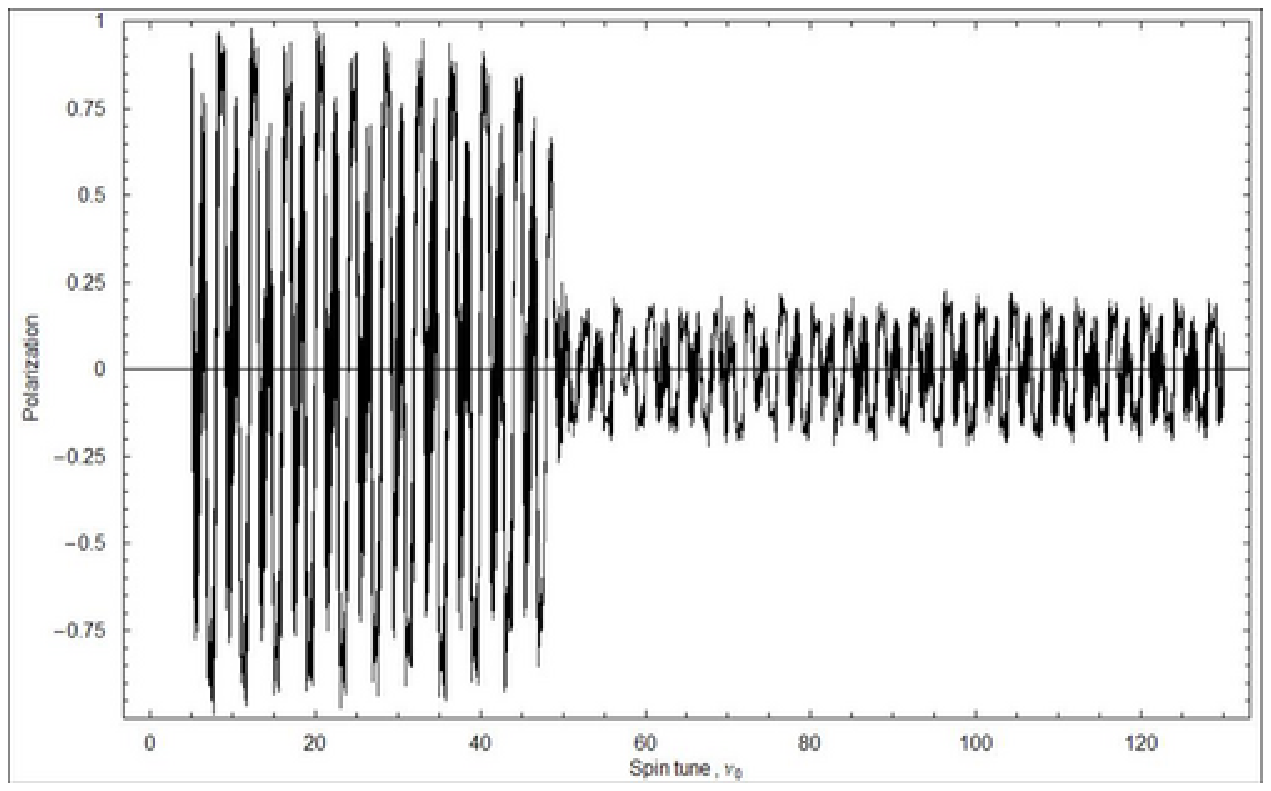} &
\includegraphics[width=0.49\textwidth] 
{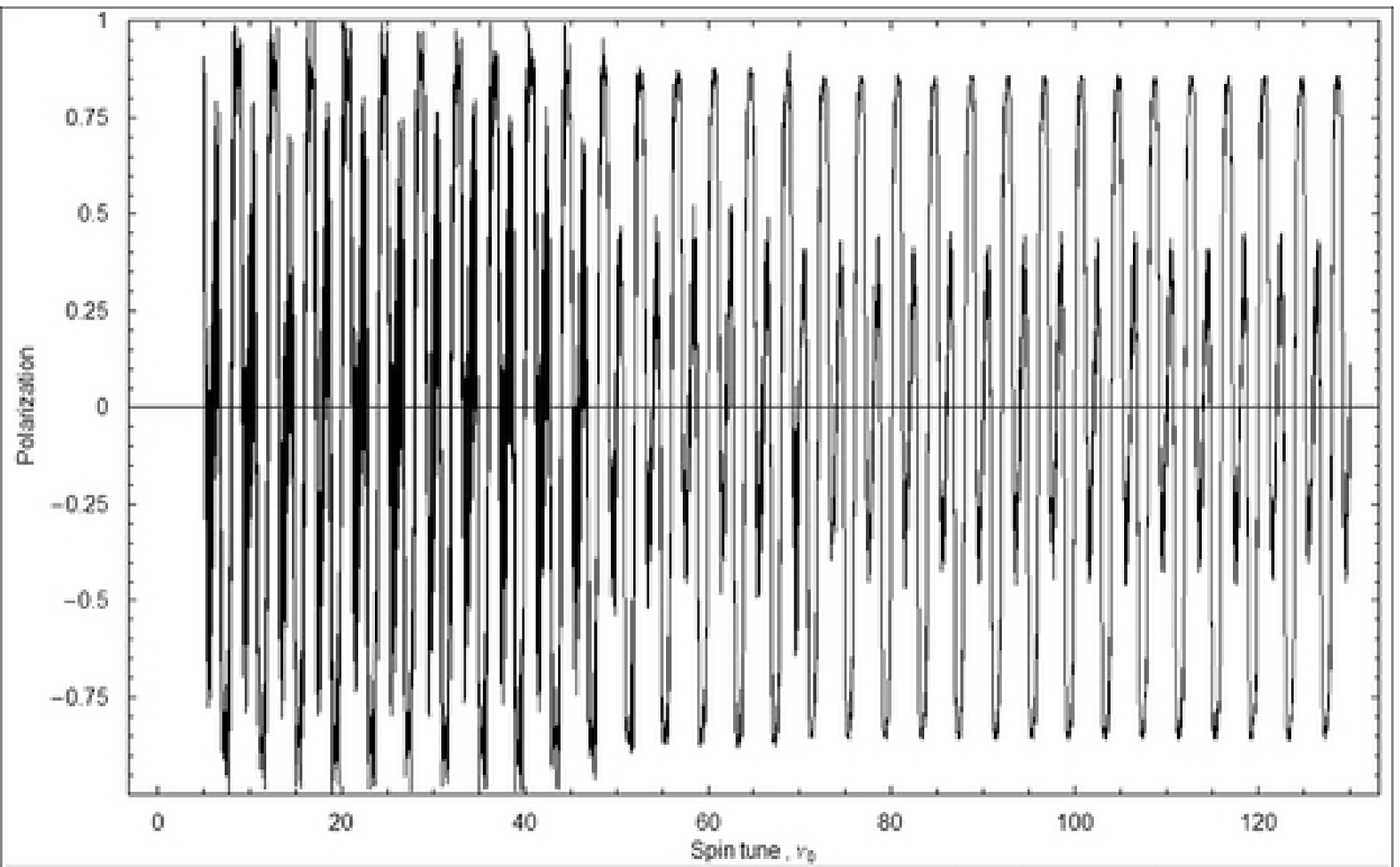} \\
\vspace*{-0.7cm}
\end{tabular}
\caption{Spin tracking of vertical polarization for 30 particles with 
two values of betatron tunes $\beta_Z=9.70$ (left) and $\beta_Z=9.94$ (right)}
\label{fig:tune}
\end{figure}

Proposed schemes using three helical partial snakes 
at U-70 can provide acceleration of polarized protons up 
to top energy E=70 GeV under conditions of conservation 
of normalized vertical emittance accepted from the booster through a whole 
acceleration cycle and alignment requirements for 
machine magnets of around $\pm 0.5$ mm. Spin tracking study proves the 
possibility to preserve polarization at a high level
while setting the prescribed vertical betatron tune $\beta_Z=9.94$
({\bf Fig.~\ref{fig:tune}}, right) in course crossing the spin resonance.

\smallskip
\subsubsection{Extraction of polarized proton beam}

The beam depolarization during extraction seems to be neglectable
because transversely (vertical) polarized beam is deflected in 
horizontal direction by vertical field. Stochastic method of beam 
slow extraction (being developed at IHEP) is seems applicable for 
polarization conservation
to extract the particles from the beam core. Expected values of 
relative polarization (in respect to polarization in main ring) 
as well as beam characteristics are presented in 
{\bf Table~\ref{tab:channel}}. Depolarization 
in the beam transfer line magnet elements is also very small due to:
\begin{itemize}
\itemnew  Relatively small $\pup$ beam emittance 
(if to compare with secondary beams);
\itemnew Opposite direction of magnetic field in 
focusing and defocusing lenses.
\end{itemize}

\begin{table}[t]
\centering
\caption {Extracted Beam characteristics at the experimental setups}
\label{tab:channel}
\begin{tabular}{|l|c|c|c|c|}
\hline
Experimental setup & SVD & SPIN & FODS & SPASCHARM \\
\hline
Size in horiz. plane, ($\sigma_x$, mm) & 1.09 & 0.58 & 1.57 & 0.22 \\
Size in vert. plane, ($\sigma_y$, mm) & 0.84 & 0.94 & 1.42 &  1.56 \\
Beam polarization, ($s_y$)  & 0.96 & $\approx 1$ & 0.99 & $\approx 1$ \\
\hline
\end{tabular}
\end{table}

\subsubsection{Polarimetry}
There should be two types of polarimeters at U70: 
\begin{enumerate}
\itemnew absolute polarimeter to determine a value 
and a sign of polarization with a high accuracy and
\itemnew less accurate but fast relative polarimeter.
\end{enumerate}
 
The both polarimeters should be based on Coulomb-Nuclear 
Interference (CNI) effect. The absolute polarimeter 
could be built with the use of a polarized jet target, 
while the relative polarimeter with the use of Carbon 
thin target. The absolute polarimeter is requred to 
calibrate the relative one. The absolute value of the 
beam polarization in the U70 averaged over 50 hours 
data taking could be measured to 5$\%$ statistical 
accuracy with the use of polarized hydrogen jet.
Relative polarimeter allowes to measure beam polarization
with the same accuracy 5\% each two minutes.  

\subsection{Proposal of the experiment SPASCHARM}

The study of spin effects in some processes would yield
information on the contribution of the spin of quarks
$\Delta\Sigma$ and gluons $\Delta$G and orbital angular 
momenta of quarks $\L_q$ and gluons $\L_g$ into the 
hadron helicity:

\begin{equation}
1/2=1/2\Delta\Sigma+L_q+\Delta G+L_g
\end{equation}

In the above sum all terms have clear physical 
interpretation, however besides the first one, they are 
gauge and frame dependent. At present only 1/3 of the 
longitudinally polarized proton spin 
is described by quark's spin. 70\% of proton spin may be 
explained by gluon and/or orbital moment contributions. 
Experiments at CERN, HERA, SLAC (polarized lepton beams) 
measured mainly quark polarization for last 15 years. 
COMPASS and HERMES are trying to measure gluon 
polarization at small $x= 0,1-0,15$. RHIC experiments 
($A_{LL}$ in direct gamma production) begin to measure 
gluon polarization at very low values of  $x$ 
($\approx 0.01)$ whereas gluon polarization should 
be measured in a whole range of $x$. 

\begin{wrapfigure}{R}{7.cm}
\mbox{\epsfig{figure=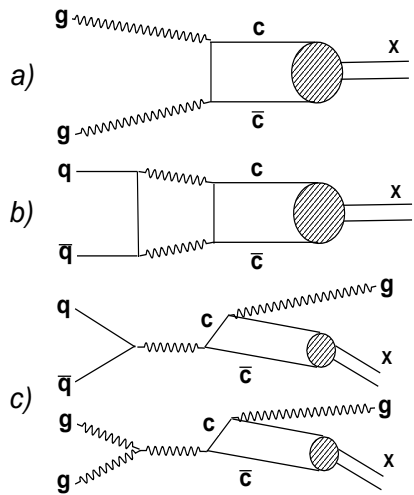,width=6cm,height=7.cm}}
\hspace*{0.3cm}
\begin{minipage}[t]{6.5cm}
\vspace*{-0.8cm}
\caption {Lowest-order parton fusion diagrams for hadronic $\chi$ production:
a) gluon fusion, b) quark-antiquark annihilation, c) color evaporation.}
\label{fig:diagram}
\end{minipage}
\end{wrapfigure}

We propose to simultaneously measure the double-SPin 
ASymmetry $A_{LL}$ in CHARM production (SPASCHARM) for 
inclusive $\chi_2$, $\chi_1$ and $J/\psi$ by utilizing 
the 70 GeV/c longitudinally 
polarized-proton beam on a longitudinally polarized target.
Our goal is to obtain besides the quark-spin information 
also the gluon-spin information from these three
processes in order to determine what portion of the 
proton spin is carried by gluons. 
One of the possible ways to measure such parton's 
polarization is a study of $\chi_c$-meson production 
with the following decay into $J/\psi$ and a
photon and then $J/\psi \to \ell^+ \ell^-$. It was shown
in~\cite{jpsipol}, that the angular distribution of the 
final photon and lepton pairs provides a direct way to 
measure the polarization of the initial quarks and gluons.
We anticipate 
obtaining significant numbers of  $\chi_2$, $\chi_1$ 
and $J/\psi$ events. This would be the world's 
first measurement of gluon-spin information and of
spin effects in charmed-particle production in 
hadron-hadron interactions.

The hadronic production of the $\chi$ states involves 
three parton fusion diagrams 
({\bf Fig.~\ref{fig:diagram}}): gluon fusion, 
light quark annihilation and color evaporation.
However, the relative contributions of each subprocess 
and even the total cross-section for charmonium 
production have proven difficult to calculate
reliably. There are fairly definite predictions
for the relative production rates of the $\chi$ states 
which may help distinguish among the models. 
The theoretical predictions for the ratios 
$\sigma (\chi_1)/ \sigma (\chi_2)=\sigma (\chi (3510))/\sigma (\chi (3555))$
are as follows \cite{ratios}: zero for gluon fusion, 
4.0 for light quark fusion and 0.6 for color evaporation. 

In this experiment the separation of $\chi_1$(3510 MeV) 
and $\chi_2$(3555 MeV) is possible. The matrix element 
for $\chi_1$ production via gluon fusion is calculated
to be zero according to the lowest-order QCD 
\cite{charmonium}. $\chi_1$ can not be produced via gluon fusion
because two spin-1 particle can not contribute to spin-1 
particle production. 
If few $\chi_1$ events are detected, 
then gluon fusion is dominant in $\chi_2$ production. 

Measurement of the $\chi_2$ production asymmetry via gluon-gluon fusion
is a cleanest way to define gluon contribution: 
\be
A_{LL}(\xf) = \hat{A}_{LL} \cdot \Delta G/G (x_1) \cdot  \Delta G/G (x_2)
\ee

\begin{wrapfigure}{R}{7.cm}
\hspace*{0.4cm}
\mbox{\epsfig{figure=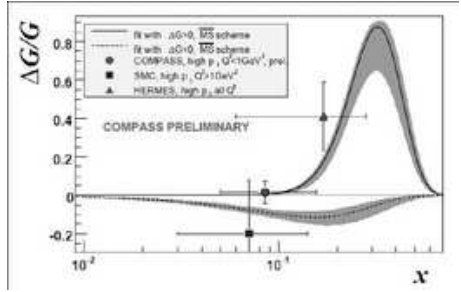,width=6.cm}}\\
\hspace*{0.3cm}
\begin{minipage}[t]{6.5cm}
\vspace*{-0.4cm}
\caption {The solution of $\Delta G/G$ from experiment COMPASS. Figure from \cite{santos}.}
\label{fig:santos}
\end{minipage}
\end{wrapfigure}

\noindent where $x_1 , x_2$ -part of proton momentum, carried out by gluons
$(M^2 (\chi_ 2) =  x_1 \cdot x_2 \cdot S$), 
$\xf$ - Feynman variable of $\chi_2$ -state,  
$\hat {A}_{LL}$-- asymmetry on parton level for two fully polarized 
gluons ($\hat {A}_{LL} =-1$ in the model of gluon-gluon fusion).
If the number of $\chi_1$ events is significant, we need
also to include other processes in the calculations to 
determine $\Delta$G/G. Note that 
$A_{LL}$($\pdup + \pdup \to \chi_1 + X$) and the 
$\chi_1 \to J/\Psi + \gamma$ decay angular 
distribution will be measured simultaneously in this case,
providing an additional input for understanding the 
production process and the value of
$\Delta G/G$ near $x=0.3$. 
Last result of COMPASS experiment indicates that this region 
is extremely important \cite{santos}, because $\Delta G/G$ has 
a maximum exactly at $x=0.3$ in one solution 
({\bf Fig.~\ref{fig:santos}}).

\begin{figure}[b]
\vspace*{-0.4cm}
\parbox[l]{0.48\hsize}{
\includegraphics[width=\hsize]{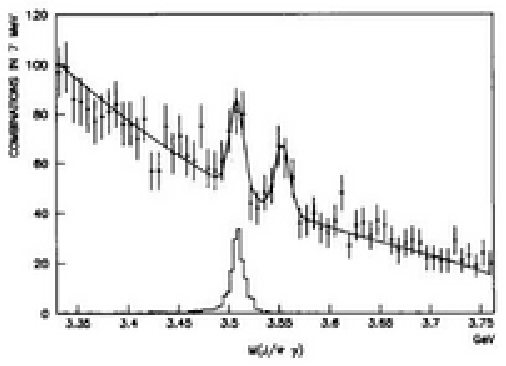}
\vspace*{-0.8cm}
\caption{$J/\Psi + \gamma$-mass spectra from the Goliath experiment}
\label{fig:goliaph}
}
\hfill
\parbox[l]{0.48\hsize}{
\includegraphics[width=\hsize,height=4.cm]{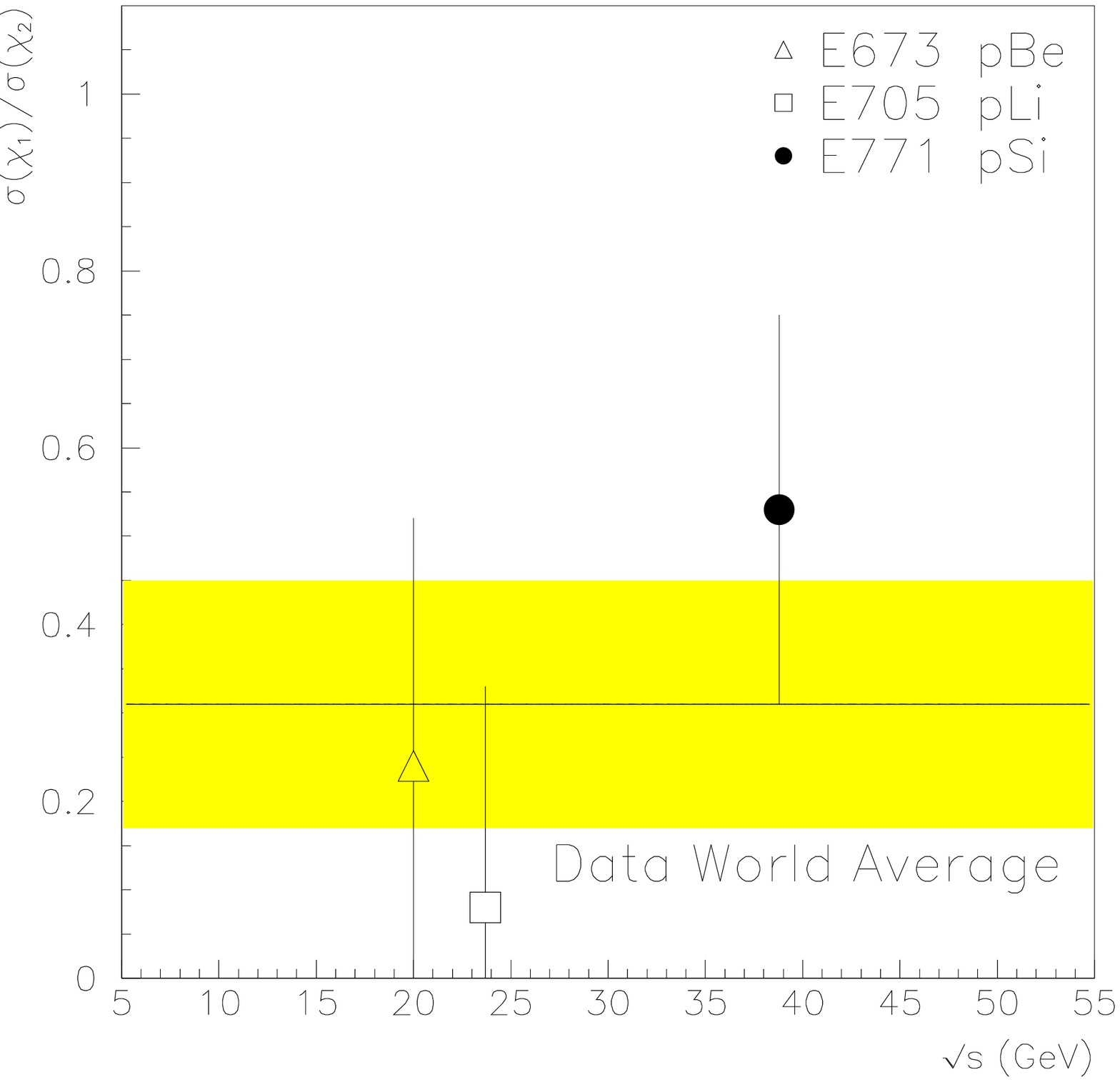}
\vspace*{-0.8cm}
\caption{World data on $\sigma (\chi_1)/\sigma (\chi_2)$ ratio}
\label{fig:crosschi}
}
\end{figure}

If all three charmonium production processes contribute, 
the measurement of $\chi_2$, $\chi_1$, and J/$\Psi$ 
become equally important. The $A_{LL}$ 
($\chi_2$, $\chi_1$, J/$\Psi$) provide a test to 
various models, which predict opposite $A_{LL}$ signs. 
The signs and magnitudes of the asymmetries will provide crucial 
information on the production mechanism(s), if the $\chi_1$ production 
at 70 GeV in pp-interactions is not negligible
compare to the $\chi_2$ production.

The study of $\chi_c$ states was carried out by Goliath and E-771
experiments. Registration of the $\gamma$-quanta from $J/\Psi$ decay 
using it's conversion to $e^+e^-$ pair allowed to distinguish between
$\chi_c$ states ({\bf Fig.~\ref{fig:goliaph}}), but significantly 
decreased the efficiency and statistics. Pure statistics did not 
allow to measure the cross-section ratio 
$\sigma (\chi_1) /\sigma (\chi_2)$ with good accuracy. 
The error of the world average value ($0.31 \pm 0.14$) is too 
high to define charmonium production mechanism 
({\bf Fig.~\ref{fig:crosschi}}).

$J/\Psi$ production at 40 GeV was studied in $\pi^-Cu$ interactions
\cite{lepton}. Total $J/\Psi$ production cross-section at 
$\xf >0$ is (980$\pm$120)~nb for Cu nucleus or $\approx$ 15~nb
per nucleon. Total $J/\Psi$ production cross-section in 
$pp$-interaction at 40 GeV is about 2~nb per nucleon, but
energy dependence of the cross-section is significant due
to the increase of the gluon fusion contribution with energy. 
The expected value of the  $J/\Psi$ production cross-section in 
$pp$-interaction at 70 GeV is 15~nb.

\subsubsection{MC-simulation}

\begin{figure}[b]
\vspace*{-0.8cm}
\centering
\begin{tabular}{cc}
\includegraphics[width=0.45\textwidth]
{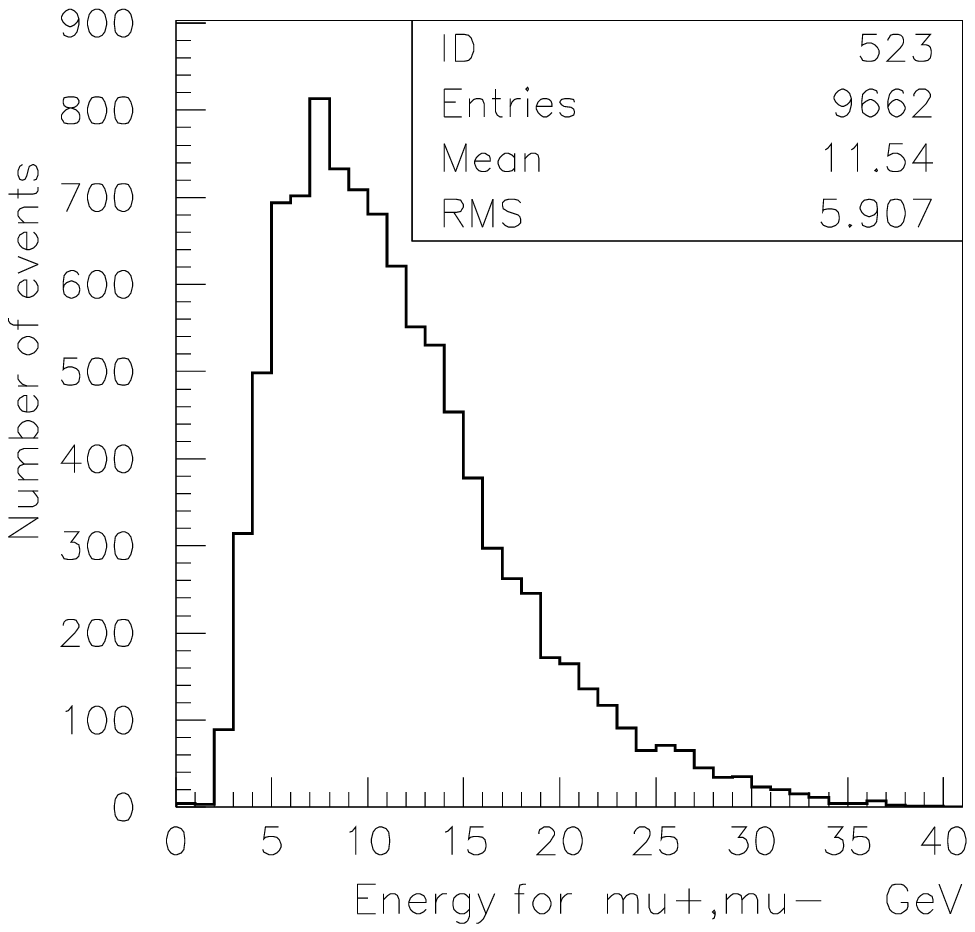} &
\includegraphics[width=0.45\textwidth]
{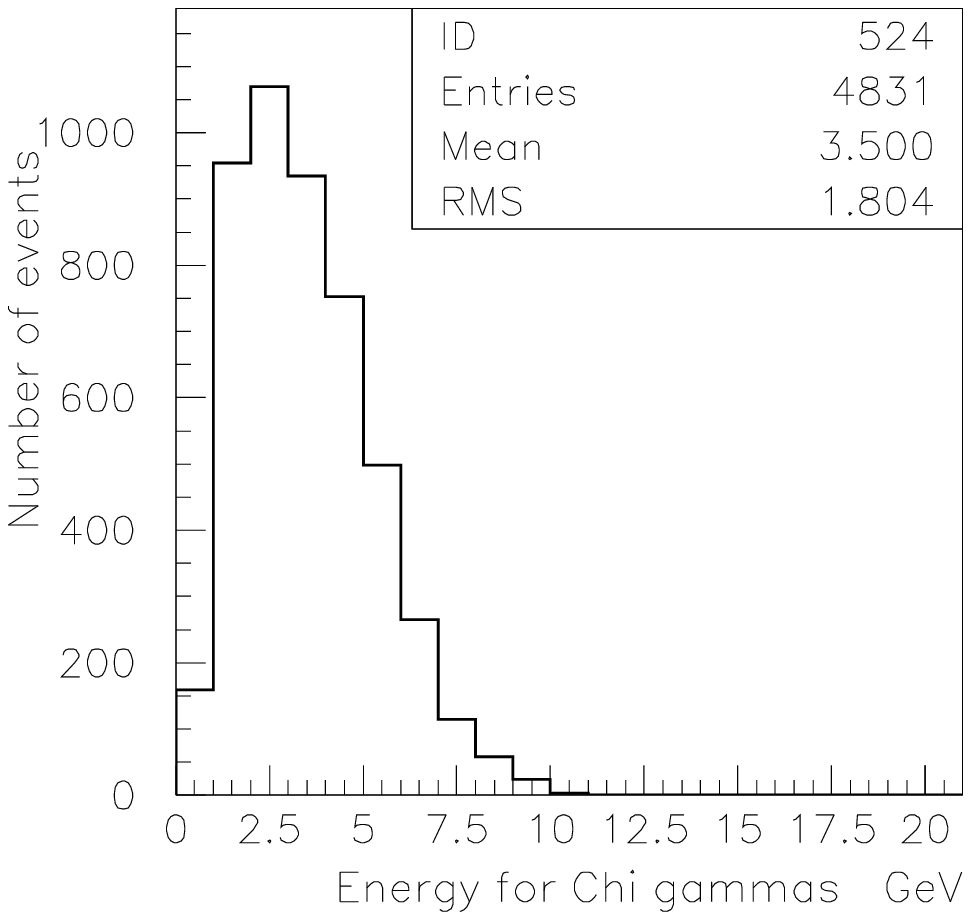} 
\\
\end{tabular}
\vspace*{-0.8cm}
\caption{Energy distribution of $\mu^+(\mu^-)$  from 
$J/\Psi$-decay (left) and $\gamma$-quanta 
from $\chi_c$-decays. }
\label{fig:energy_reg}
\end{figure}  

The main experimental challenge is to separate two 
close narrow states --- $\chi_1$(3510~MeV) and 
$\chi_2$(3555~MeV). Mass resolution must be better, than
their 1\% mass difference. In order to study the 
requirements for the experimental equipment and to
estimate the expected statistics MC-simulation of the 
processes was carried out at following conditions:
\begin{itemize}
\itemnew PYTHIA was used as a standard generator. 
\itemnew Only gluon-gluon fusion and color evaporation 
diagrams for the $\chi_c$ production were used for 
reconstruction study (including quark-antiquark annihilation 
diagram does not change mass spectra). 
\itemnew $\chi_c$ - states ($\chi_0(3410), \chi_1(3510)$ 
and $\chi_2(3555)$) decays to $J/\Psi + \gamma$  were studied. 
\itemnew $J/\Psi$-decay was studied only in muon mode.
\itemnew MC was carried out for several values of the 
charge particle's momentum resolution and for the two 
values of the energy resolution of an electromagnetic 
calorimeter (ECal). The accuracy of the primary vertex 
reconstruction in beam direction is $\sigma_z = 10$~mm. 
\itemnew The value of charge particle track momentum resolution 
is presented for 10~GeV particle. 
\itemnew The influence of kinematic $1C$-fit for $J/\Psi$  
($J/Psi$- mass is fixed, 3-momentum components are fitted) 
to $\chi_c$  states mass resolution and the improvement of 
the separation was studied.
\end{itemize}

\begin{figure}[b]
\parbox[l]{0.48\hsize}{
\vspace*{-0.8cm}
\includegraphics[width=\hsize]{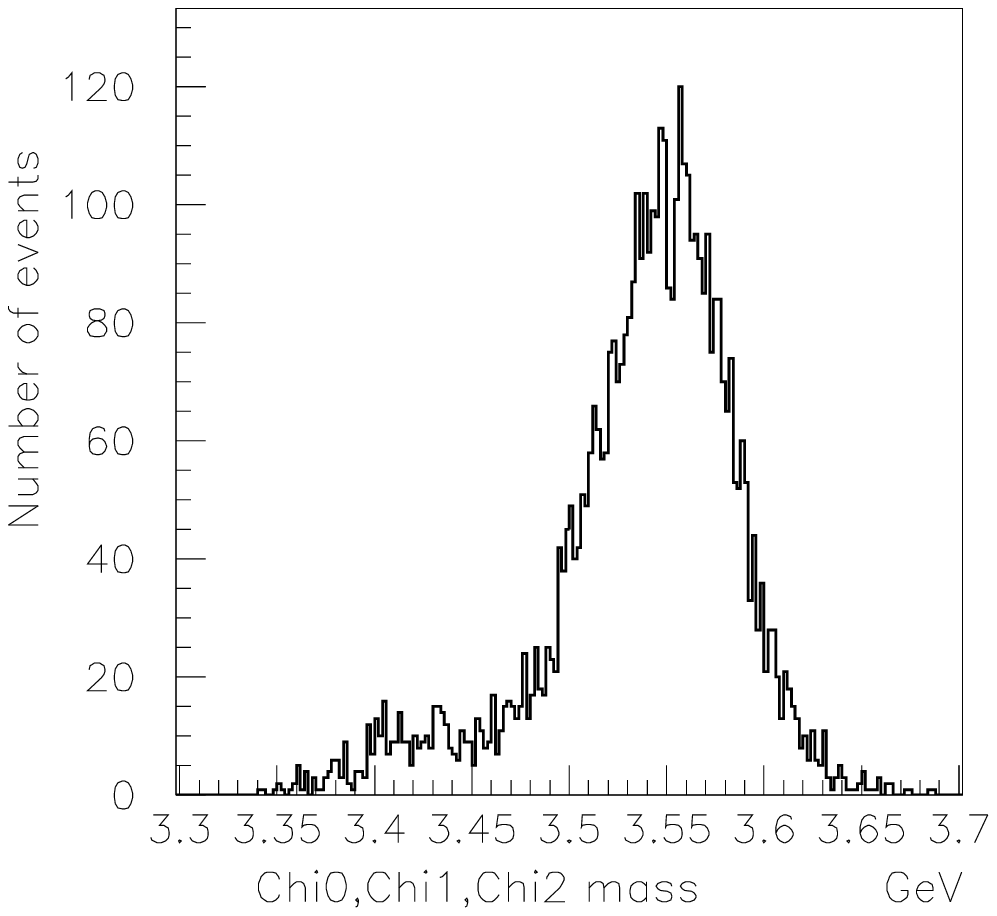}
\vspace*{-1.3cm}
\caption{Reconstruction of $\chi_c$-mass {\bf without} 1C 
fit. Track momentum 
resolution $\delta p/p=0.3$\%. EMC energy resolution 
$\sigma (E)/E= 12\%/\sqrt{E}$.}
\label{fig:minaev_eff1}
} 
\hfill
\parbox[l]{0.48\hsize}{
\vspace*{-0.8cm}
\includegraphics[width=\hsize]{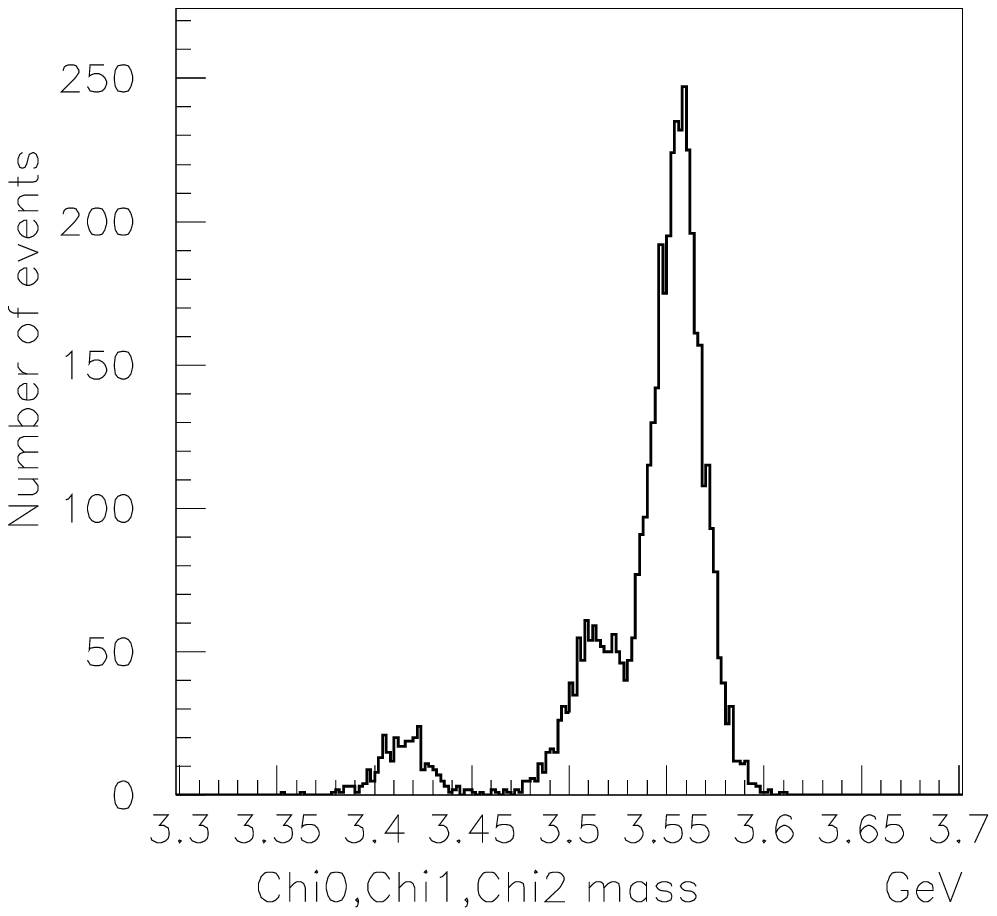}
\vspace*{-1.3cm}
\caption{Reconstruction of $\chi_c$-mass {\bf without} 1C fit. Track momentum 
resolution $\delta p/p=0.4$\%. EMC energy resolution 
$\sigma (E)/E= 2.5\%/\sqrt{E}$.}
\label{fig:minaev_eff2}
}
\parbox[l]{0.48\hsize}{
\vspace*{-0.3cm}
\includegraphics[width=\hsize]{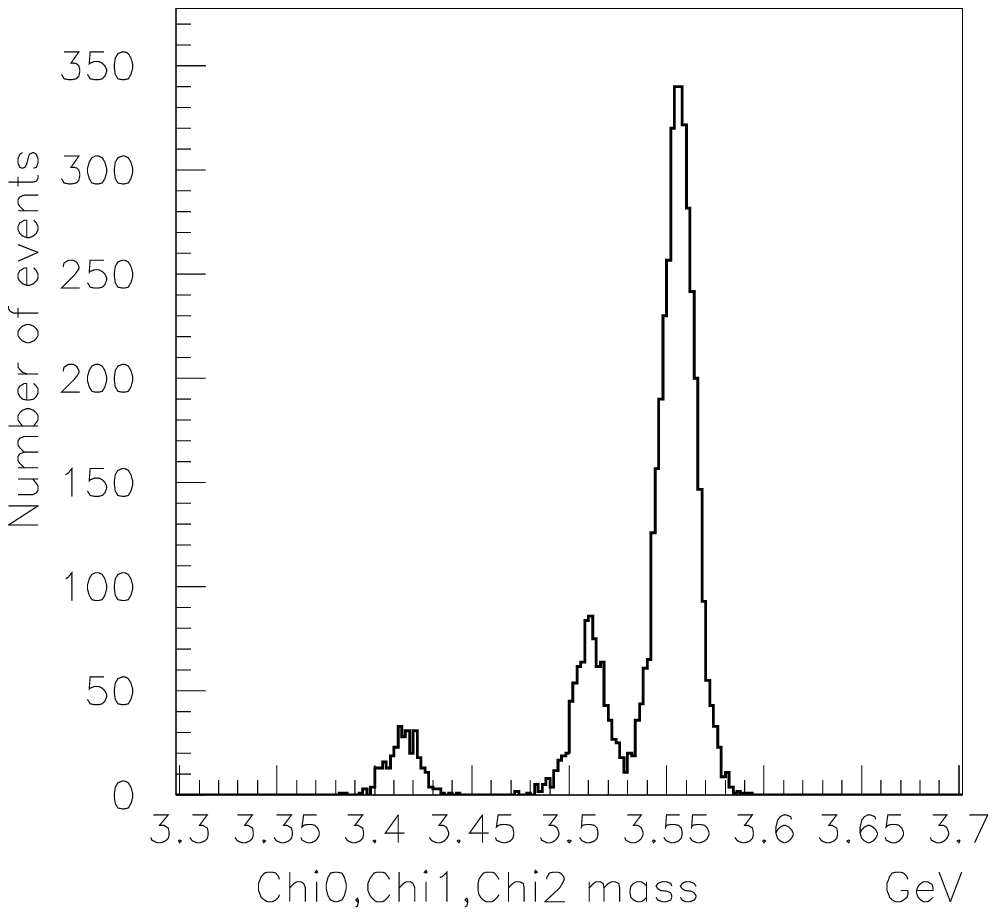}
\vspace*{-0.8cm}
\caption{
Reconstruction of $\chi_c$-mass {\bf with} the 
use of $1C$ fit on $J/\Psi$-mass. 
$\delta p/p=0.4$\%. $\sigma (E)/E= 2.5\%/\sqrt{E}$.}
\label{fig:minaev_eff3}
}
\hfill
\parbox[l]{0.48\hsize}{
\vspace*{-0.3cm}
\includegraphics[width=\hsize]{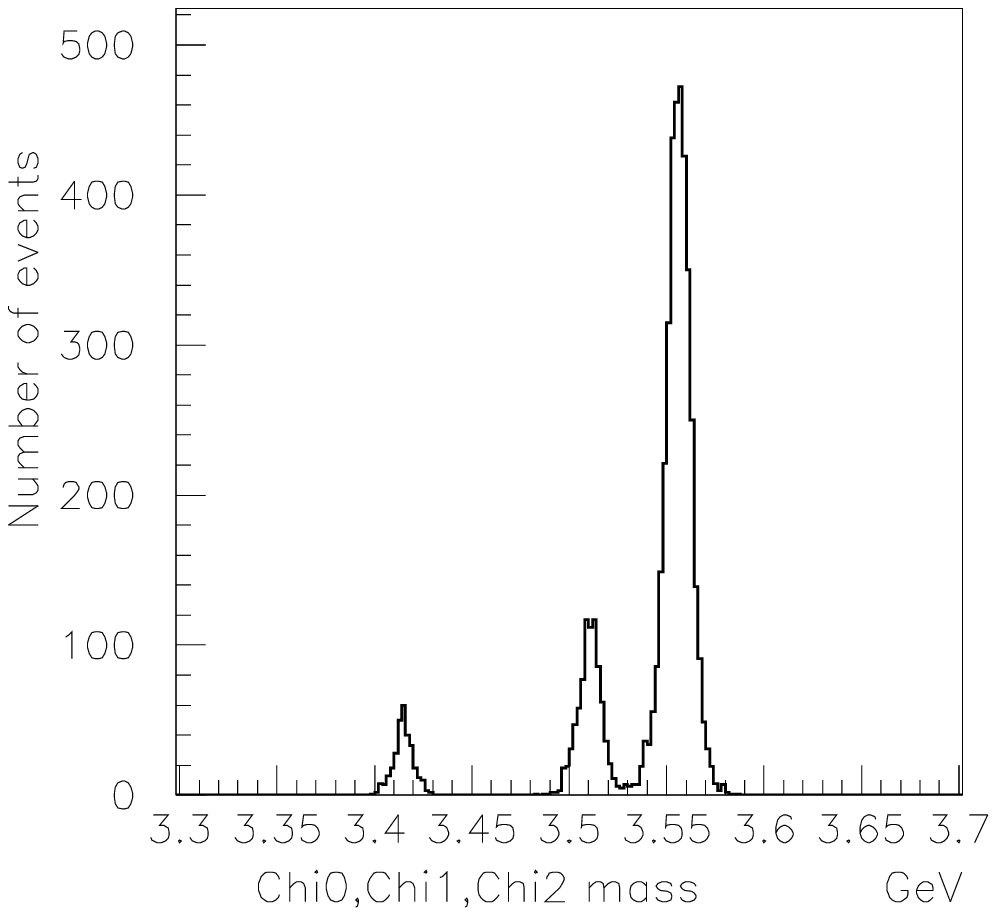}
\vspace*{-0.8cm}
\caption{
Reconstruction of $\chi_c$-mass with 
$\mu^+(\mu^-) $ momentum from  PYTHIA. 
$\sigma (E)/E= 2.5\%/\sqrt{E}$.}
\label{fig:minaev_eff4}
}
\end{figure}

Average energy of $\mu^+(\mu^-)$-mesons registered 
from $J/\Psi$ decay is 11.5~GeV and average energy of 
registered $\gamma$-quanta from $\chi_c$-decay is 
3.5~GeV ({\bf Fig.~\ref{fig:energy_reg}}). 
According to the decay kinematics of $\chi_2$,
the $\gamma$'s are effectively detected at very forward 
angle (up to 150~mrad). A calorimeter for detection of 
these $\gamma$'s must have good energy resolution.
$\chi_c$ states can not be disentangled by lead-glass 
electromagnetic calorimeter with the resolution 
$\sigma (E)/E = 12.5\%/\sqrt{E}$ (see 
{\bf Fig.~\ref{fig:minaev_eff1}}) even with perfect 
momentum resolution  $\delta p/p=0.3$\%. The distribution
does not change when introducing $1C$ fit on $J/\Psi$-mass.

$\chi_c$ peaks are clearly seen when using the calorimeter 
with resolution $\sigma (E)/E = 2.5\%/\sqrt{E}$ (see 
{\bf Fig.~\ref{fig:minaev_eff2}}) even with slightly 
worse momentum resolution  $\delta p/p=0.4$\% and 
without $1C$ fit procedure on $J/\Psi$-mass. 
Introducing $1C$ fit procedure on $J/\Psi$-mass improves the
distribution ({\bf Fig.~\ref{fig:minaev_eff3}}). The
mass spectra with generated charge particle momenta is 
also presented on {\bf Fig.~\ref{fig:minaev_eff4}}.

\subsubsection{Experimental Setup}

\begin{figure}[b]
\centering
\includegraphics[width=0.95\textwidth]
{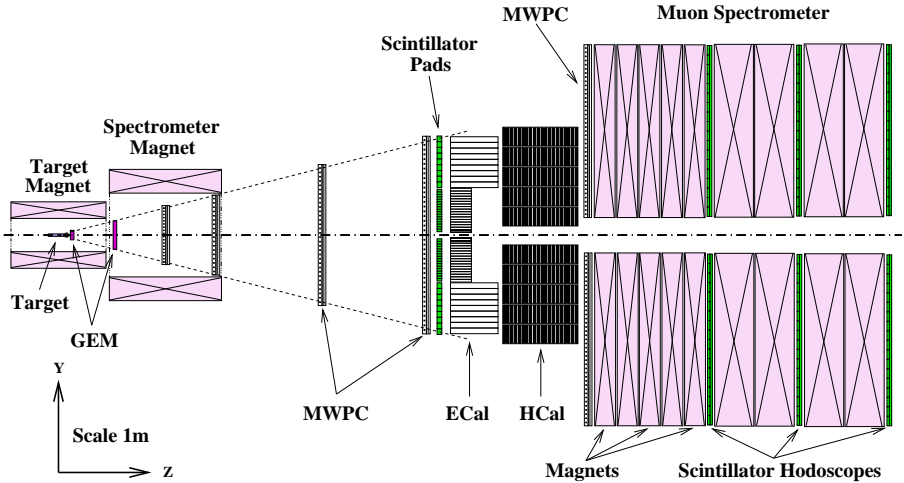} 
\vspace*{-0.8cm}
\caption{SpAsCharm Experimental Setup}
\label{fig:spascharm_setup}
\end{figure}  

MC study proved the possibility to separate $\chi_c$-states
and provided following requirements for the experimental setup: 
\begin{itemize}
\itemnew Possibility to work with $5 \cdot 10^6$ 
interactions per cycle (3 sec table, duty factor 1/3, 
beam intensity $5 \cdot 10^7$ p/cycle).
\itemnew Large acceptance for Charmonium and Drell-Yan 
processes detection;
\itemnew EMC energy resolution $\sigma (E)/E=2.5\%/\sqrt{E}$ 
at least at the central region, required for $\chi_c$ 
separation.
\itemnew Charge particle momentum resolution 
$\delta p/p=0.4\%$ at 10~GeV. 
\itemnew Good separation of electrons and muons from hadrons, 
(charm production is at the level of $10^{-7}$-$10^{-8}$ 
per interaction in the target).
\itemnew Very fast DAQ to write information with the rate 
up to 100 Mb/sec (to study different inclusive processes 
with large cross-section). 
\end{itemize}

The proposed schema of the experimental setup 
({\bf Fig.~\ref{fig:spascharm_setup}})
consists of seven different systems.

Fast scintillation counters and fiber hodoscopes will 
be used for beam trigger and coordinate reconstruction. 
Such type of fiber hodoscopes with the diameter of 0.28 mm 
was developed for DIRAC experiment. Light response 
is 10-12 photoelectrons for each fiber. 

Transversely and longitudinally polarized targets
are required for double-spin asymmetry measurements. The 
best solution is $NH_3$ target with the best figure of
merit for polarized hydrogen targets.

Magnet spectrometer will consist of the magnet 
with the aperture 1.2 (horizontal) $\times $ 1 (vertical) 
$\times $ 1~m$^3$ and field 15~kGs.  
Two or even three triple-GEM detectors with coordinate 
resolution better than 70 microns (or similar) will be used 
as the tracking detectors before the magnet. 
Three stations of MWPC or mDC will be used 
for the tracking detectors after spectrometer magnet. Such
schema allows to obtain momentum resolution better than
$\delta p/p=0.04 \cdot P \cdot 10^{-2}$c/GeV without taking 
into account multiple scattering.

Electromagnetic calorimeter for $\gamma$-quanta
registration will consist of two parts. The central part 
($\Theta_{lab}$ from 10 mrad to 125 mrad)
of the calorimeter system consists of 1216 blocks of 
lead tungstate (2.8 $\times$ 2.8 $\times$ 22 cm$^3$ per 
each block) to ensure good energy resolution of the 
$\gamma$ detection. This is an array of 35$\times $35 blocks
with a hole of 3$\times $3 blocks in the center for 
non-interacted beam. The properties of lead tungstate (PWO)
calorimeters have been extensively studied at IHEP over last 
several years \cite{nim8}.  The energy resolution of ~2$\%$ at
$E = 1$~GeV has been measured. The 2028 lead-glass or 
lead-scintillator counters (3.81 $\times$ 3.81 $\times$ 45 cm$^3$ 
per each block) cover a large area ($\Theta_{lab}$ from 
125 mrad to 250 mrad) and will be used mainly for separation
of electrons and hadrons together with tracking system and
hadron calorimeter. This is an array of 52x52 blocks with 
a hole of 26$\times $26 in the center to accept the lead 
tungstate blocks.

The scintillator-pad trigger hodoscope containing 
100 pads segmented mosaic structure before ECal will 
be used  for the trigger.

Compensated lead-scintillator hadron calorimeter \cite{hcal}
will be assembled from $10 \times 10$~cm$^2$ modules (6.5
nuclear length). 

A muon detector will be upgraded on base of existing 
muon spectrometer of JINR-IHEP neutrino detector. The block
of proportional chambers will be installed before muon detector
to coincide muon tracks with charged ones in tracking system.
\smallskip

\subsubsection{Efficiency and expected accuracy}

\begin{figure}[t]
\parbox[l]{0.48\hsize}{
\includegraphics[width=\hsize,height=4.cm]{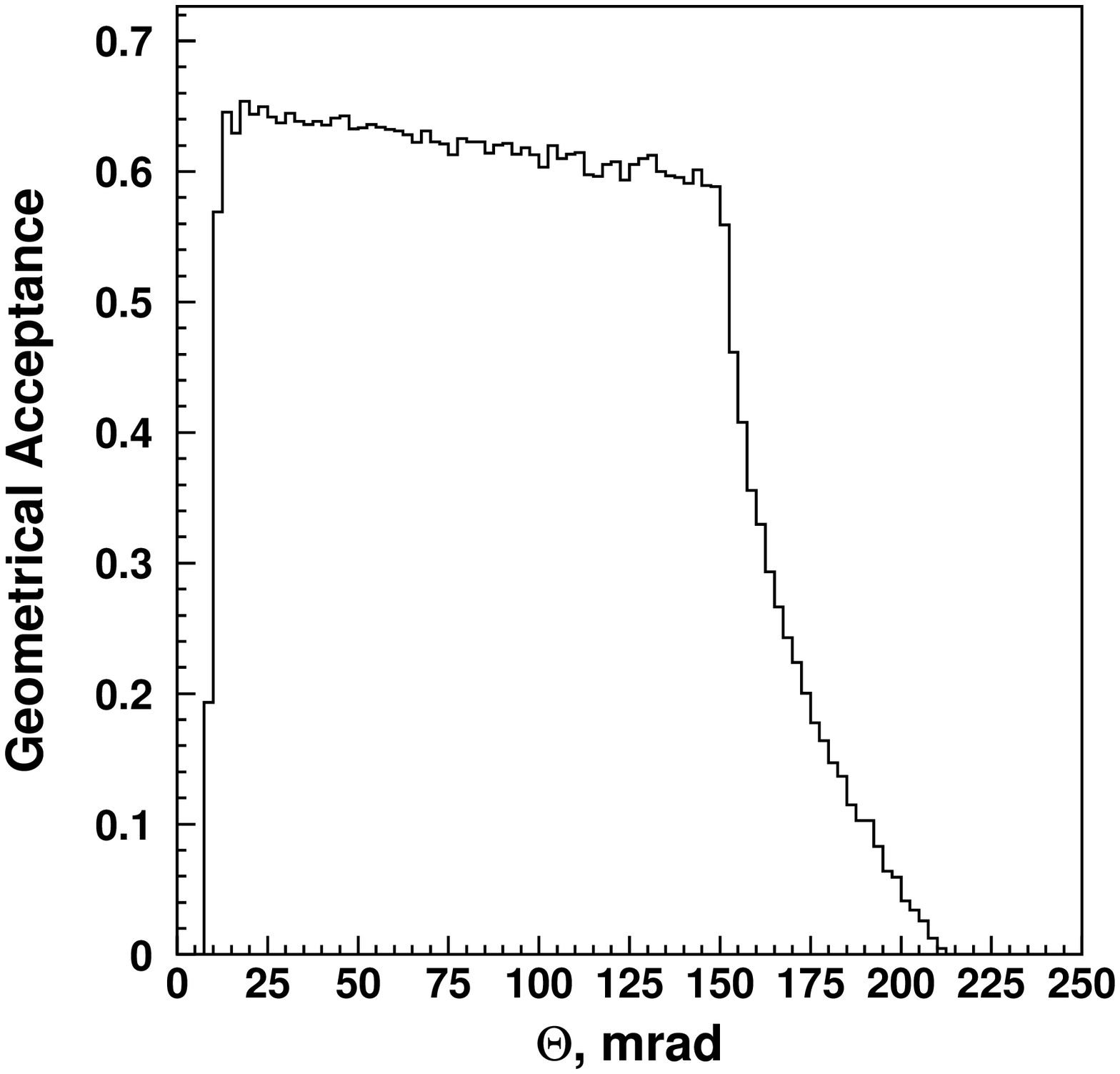}
\vspace*{-1.cm}
\caption{Geometrical Acceptance for $\chi_2$ at $\xf>0$, 
$\gamma$-quanta is detected in PWO ($10<\Theta<125$ mrad,  
$\mu^+(e^+)$ and $\mu^-(e-)$) from $J/\Psi$ decay are 
detected in the angle range $10<\Theta<250$}
\label{fig:geom_accept}
}
\hfill
\parbox[r]{0.48\hsize}{
\vspace*{-1.2cm}
\includegraphics[width=\hsize,height=4.cm]{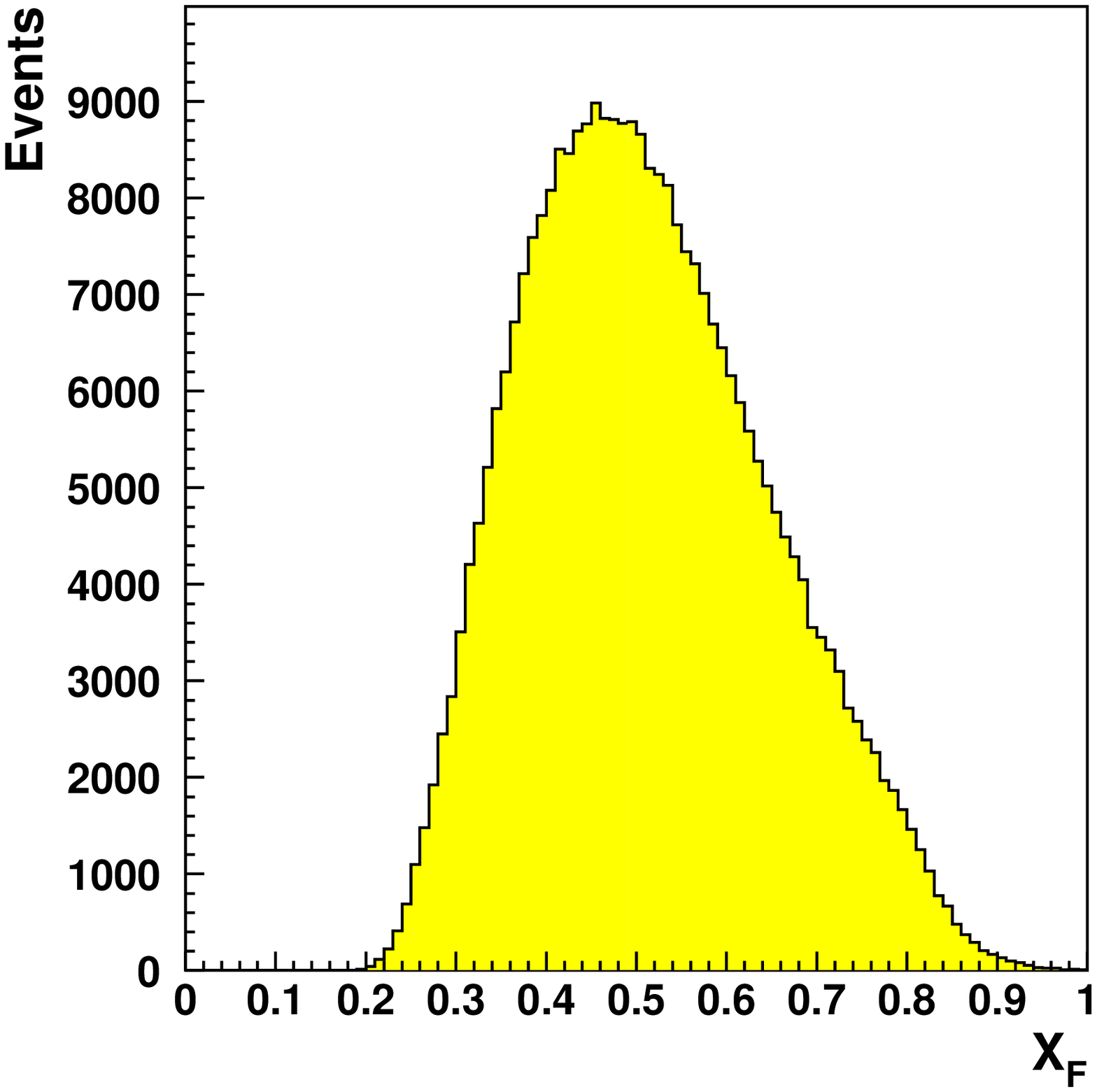}
\vspace*{-1.cm}
\caption{Kinematic $\xf$ region of detected $\chi_2$ state}
\label{fig:chi2_xf}
}
\end{figure}

The geometry of the experimental setup was chosen to 
achieve the maximal efficiency in $J/\Psi$ and $\chi_c$ 
registration. Geometry acceptance is 61\% for $J/\Psi$ 
registration in lepton decay mode and Drell-Yan and 
31\%for $\chi_c$-states ({\bf Fig.~\ref{fig:geom_accept}}). 
Kinematic $\xf$ region of detected $\chi_2$ is presented on 
{\bf Fig.~\ref{fig:chi2_xf}}. The conservative estimation 
of $J/\Psi$ event rate was done at following conditions:
\begin{itemize}
\itemnew existing polarized propane-diol 20~cm long target;
\itemnew geometrical acceptance is 60\%;
\itemnew trigger, reconstruction and accelerator efficiencies
are 0.9, 0.6 and 0.8 correspondingly.
\itemnew beam intensity is $5 \cdot 10^6$ $\pi^-$/cycle at 40~GeV
and $5 \cdot 10^7$ $p$/cycle at 70~GeV.
\itemnew $J/\Psi$ registration in two lepton modes.
\end{itemize}

$J/\Psi$ will be produced with the probability 
$2.42 \cdot 10^{-9}$ for incident particle or 4-5 events per
hour. The integral number of registered $J/\Psi$ is 
4000 for pion beam for 40-day of statistics data taking. 
Single spin $A_N$ statistic error is 14\%.
The number of $J/\Psi$ registered in $pp$-interactions is
ten times larger for the same time. The accuracy of 
single-spin asymmetry is 4.5\% for $J/\Psi$ and 
12-20\% for $\chi_1/\chi_2$-states. The statistics may be 
increased by factor of two by using $NH_3$ target and by
additional factor of two increasing extracted beam intensity.

Statistics may be increased for $J/\Psi$  as well as for 
$\chi_1/\chi_2$-states studying $J/\Psi$ production
also in hadron modes. Geometrical efficiency (with branching) 
for two lepton modes is 7.2\%, for three hadron modes 
2.5\%. We conservatively estimated ECal energy resolution 
$\sigma (E)/E = 8\%/\sqrt{E}$ for hadron modes, assuming 
that at least one gamma from $\pi^0$-decay is outside 
Ecal central region. 

\subsubsection{Experiment schedule}

Two main stages of the experiment are devoted to different 
main tasks. Stage 1 (with "old" polarized target and without 
polarized beam) is devoted to single-spin asymmetry 
measurements of $J/\psi$ and $\chi_c$ inclusive production 
using pion and unpolarized proton beam. 
The cross-section ratio for $\chi_1/\chi_2$ production will 
be measured to determine the mechanism of charmonium 
production. 

The main goal of stage 2 (with longitudinally polarized beam 
and target) is to measure double-spin asymmetry $A_{LL}$ in 
charmonium production to study $\Delta G/G(x)$. 
Also $A_{NN}$ will be measured for Drell-Yan pairs to study 
transversity $h(x)$. Simultaneously $A_{NN}$ and $A_N$ 
in $J/\Psi$, $\chi_c$ production will be studied.  
SPASCHARM experimental setup is universal and will be used to 
study charmonium production using heavy ion beam. 

SPASCHARM construction time is 6 years (beginning 2007), but 
we are discussing the possibility to start spin studies 
at 2010 before the all experimental sub-systems will be ready. 
The possible physics is a polarization study of light and 
strange resonances in inclusive and/or exclusive reactions.
Particle identification is required for this stage. 
Existing calorimeters, magnet. muon detector and cherenkov 
counters may be used after small modernization. 
New tracking system (MWPC or mDC chambers, GEM detectors 
or similar) and new fast electronics and DAQ required.  

\section{Conclusion}
The investigation of spin properties of the matter and 
namely hadron spin structure and the role of the spin in 
hadron reactions is studied at Protvino. Unexpected 
large spin asymmetry in neutral meson inclusive production
have stimulated many theoretical models. 

Precision measurements using a polarized proton beam are
required to distinguish between the models. The acceleration 
of the polarized proton beam at U-70 accelerator will give us the
possibility to investigate nucleon spin structure. Proposal of 
new experiment to measure double-spin asymmetry in charm production 
is a cleanest way to study gluon contribution to nucleon spin. 
Spin effects in many other reactions will be measured 
simultaneously. The proposed experiment will be complementary 
to existing experiments.

\bigskip

{\small The work was partially supported by RFBR grant 06-02-16119}.

\end{document}